\documentclass[sigconf]{acmart}
\usepackage{booktabs} 
\usepackage{multirow}
\usepackage{color}
\usepackage{subfigure} 
\usepackage{array}
\usepackage{breqn}
\usepackage{{inputenc}}
\usepackage{balance}
\usepackage{multirow,tabularx}
\usepackage{longtable}
\usepackage{booktabs,longtable}
\usepackage{hhline}
\usepackage[flushleft]{threeparttable}
\usepackage{algorithm}
\usepackage{algorithmic}
\usepackage{epsfig}
\usepackage{graphicx}
\usepackage{epstopdf}
\usepackage{thmtools}
\usepackage{enumitem}
\usepackage{verbatim}
\usepackage{amsfonts}
\usepackage{hyperref}
\hypersetup{colorlinks=false,linkcolor=blue,urlcolor=blue,citecolor=red}
\newcommand{\Mat}[1]{\mathbf{#1}}

\newcommand{\Set}[1]{\mathcal{#1}}

\newcommand{\ie}{\emph{i.e., }}

\theoremstyle{definition}

\def\BibTeX{{\rm B\kern-.05em{\sc i\kern-.025em b}\kern-.08emT\kern-.1667em\lower.7ex\hbox{E}\kern-.125emX}}

\copyrightyear{2020} \acmYear{2020} \setcopyright{acmlicensed}\acmConference[SIGIR '20]{Proceedings of the 43rd International ACM SIGIR Conference on Research and Development in Information Retrieval}{July 25--30, 2020}{Virtual Event, China} \acmBooktitle{Proceedings of the 43rd International ACM SIGIR Conference on Research and Development in Information Retrieval (SIGIR '20), July 25--30, 2020, Virtual Event, China} \acmPrice{15.00} \acmDOI{10.1145/3397271.3401063} \acmISBN{978-1-4503-8016-4/20/07}

\begin{document}
\fancyhead{}

\title{LightGCN: Simplifying and Powering Graph Convolution Network for Recommendation}

\author{Xiangnan He}
\affiliation{
\institution{University of Science and Technology of China}
}\email{xiangnanhe@gmail.com}

\author{Kuan Deng}
\affiliation{
\institution{University of Science and Technology of China}
}\email{dengkuan@mail.ustc.edu.cn}

\author{Xiang Wang}
\affiliation{
\institution{National University of Singapore}
}\email{xiangwang@u.nus.edu}

\author{Yan Li}
\affiliation{
\institution{Beijing Kuaishou Technology \\Co., Ltd.}
}\email{liyan@kuaishou.com}

\author{Yongdong Zhang}
\affiliation{
\institution{University of Science and Technology of China}
}\email{zhyd73@ustc.edu.cn}

\author{Meng Wang}
\authornote{Meng Wang is the corresponding author.}
\affiliation{
\institution{Hefei University of Technology}
}\email{eric.mengwang@gmail.com
}

\setcounter{page}{1}
\pagenumbering{Roman}

\begin{abstract}
Graph Convolution Network (GCN) has become new state-of-the-art for collaborative filtering. Nevertheless, the reasons of its effectiveness for recommendation are not well understood. 
Existing work that adapts GCN to recommendation lacks thorough ablation analyses on GCN, which is originally designed for graph classification tasks and equipped with many neural network operations. 
However, we empirically find that the two most common designs in GCNs --- feature transformation and nonlinear activation --- contribute little to the performance of collaborative filtering. Even worse, including them adds to the difficulty of training and degrades recommendation performance. 

In this work, we aim to simplify the design of GCN to make it more concise and appropriate for recommendation. We propose a new model named LightGCN, including only the most essential component in GCN --- neighborhood aggregation --- for collaborative filtering. Specifically, LightGCN learns user and item embeddings by linearly propagating them on the user-item interaction graph, and uses the weighted sum of the embeddings learned at all layers as the final embedding. 
Such simple, linear, and neat model is much easier to implement and train, exhibiting substantial improvements (about 16.0\% relative improvement on average) over Neural Graph Collaborative Filtering (NGCF) --- a state-of-the-art GCN-based recommender model --- under exactly the same experimental setting. Further analyses are provided towards the rationality of the simple LightGCN from both analytical and empirical perspectives. Our implementations are available in both TensorFlow\footnote{\url{https://github.com/kuandeng/LightGCN}} and PyTorch\footnote{\url{https://github.com/gusye1234/pytorch-light-gcn}}. 

\end{abstract}
%
%

\vspace{-5pt}
\maketitle

\section{Introduction}
To alleviate information overload on the web, recommender system has been widely deployed to perform personalized information filtering~\cite{PinSage,YoutubeRS, PeterRec}.
The core of recommender system is to predict whether a user will interact with an item, e.g., click, rate, purchase, among other forms of interactions. As such, collaborative filtering (CF), which focuses on exploiting the past user-item interactions to achieve the prediction, remains to be a fundamental task towards effective personalized recommendation~\cite{NCF,VACF,NGCF,CMN}. 

The most common paradigm for CF is to learn latent features (a.k.a. embedding) to represent a user and an item, and perform prediction based on the embedding vectors~\cite{NCF,DBLP:conf/www/ChengDZK18}.
Matrix factorization is an early such model, which directly projects the single ID of a user to her embedding~\cite{MF}.  Later on, several research find that augmenting user ID with the her interaction history as the input can improve the quality of embedding. For example, SVD++~\cite{SVD++} demonstrates the benefits of user interaction history in predicting user numerical ratings, and Neural Attentive Item Similarity (NAIS)~\cite{NAIS} differentiates the importance of items in the interaction history and shows improvements in predicting item ranking. 
In view of user-item interaction graph, these improvements can be seen as coming from using the subgraph structure of a user --- more specifically, her one-hop neighbors --- to improve the embedding learning. 

To deepen the use of subgraph structure with high-hop neighbors, Wang et al.~\cite{NGCF} recently proposes NGCF and achieves state-of-the-art performance for CF. It takes inspiration from the Graph Convolution Network (GCN)~\cite{GCN,GraphSAGE}, following the same propagation rule to refine embeddings: feature transformation, neighborhood aggregation, and nonlinear activation. Although NGCF has shown promising results, we argue that its designs are rather heavy and burdensome --- many operations are directly inherited from GCN without justification. As a result, they are not necessarily useful for the CF task. To be specific, GCN is originally proposed for node classification on attributed graph, where each node has rich attributes as input features; whereas in user-item interaction graph for CF, each node (user or item) is only described by a one-hot ID, which has no concrete semantics besides being an identifier. In such a case, given the ID embedding as the input, performing multiple layers of nonlinear feature transformation --- which is the key to the success of modern neural networks~\cite{ResNet} --- will bring no benefits, but negatively increases the difficulty for model training. 

To validate our thoughts, we perform extensive ablation studies on NGCF. With rigorous controlled experiments (on the same data splits and evaluation protocol), we draw the conclusion that the two operations inherited from GCN --- feature transformation and nonlinear activation --- has no contribution on NGCF's effectiveness. Even more surprising, removing them leads to significant accuracy improvements. This reflects the issues of adding operations that are useless for the target task in graph neural network, which not only brings no benefits, but rather degrades model effectiveness. Motivated by these empirical findings, we present a new model named LightGCN, including the most essential component of GCN --- neighborhood aggregation --- for collaborative filtering. Specifically, after associating each user (item) with an ID embedding, we propagate the embeddings on the user-item interaction graph to refine them. We then combine the embeddings learned at different propagation layers with a weighted sum to obtain the final embedding for prediction. The whole model is simple and elegant, which not only is easier to train, but also achieves better empirical performance than NGCF and other state-of-the-art methods like Mult-VAE~\cite{VACF}. 

To summarize, this work makes the following main contributions:
\begin{itemize}
    \item We empirically show that two common designs in GCN, feature transformation and nonlinear activation, have no positive effect on the effectiveness of collaborative filtering. 
    \item We propose LightGCN, which largely simplifies the model design by including only the most essential components in GCN for recommendation.
    \item We empirically compare LightGCN with NGCF by following the same setting and demonstrate substantial improvements. In-depth analyses are provided towards the rationality of LightGCN from both technical and empirical perspectives. 
\end{itemize}

\section{Preliminaries}\label{sec:preliminaries}

We first introduce NGCF~\cite{NGCF}, a representative and state-of-the-art GCN model for recommendation. We then perform ablation studies on NGCF to judge the usefulness of each operation in NGCF. The novel contribution of this section is to show that the two common designs in GCNs, feature transformation and nonlinear activation, have no positive effect on collaborative filtering. 

\subsection{NGCF Brief}\label{ss:ngcf}
In the initial step, each user and item is associated with an ID embedding. Let $\textbf{e}_u^{(0)}$ denote the ID embedding of user $u$ and $\textbf{e}_i^{(0)}$ denote the ID embedding of item $i$. Then NGCF leverages the user-item interaction graph to propagate embeddings as:
\begin{equation}\label{eq:NGCF}
\begin{aligned}
    \textbf{e}_{u}^{(k+1)} &= \sigma \Big(\textbf{W}_1 \textbf{e}_u^{(k)} + \sum_{i\in \mathcal{N}_u} \frac{1}{\sqrt{|\Set{N}_{u}||\Set{N}_{i}|}}(\Mat{W}_{1}\Mat{e}_{i}^{(k)} + \Mat{W}_{2}(\Mat{e}_{i}^{(k)}\odot\Mat{e}_{u}^{(k)}))\Big), \\
    \textbf{e}_{i}^{(k+1)} &= \sigma \Big(\textbf{W}_1 \textbf{e}_i^{(k)} + \sum_{u\in \mathcal{N}_i} \frac{1}{\sqrt{|\Set{N}_{u}||\Set{N}_{i}|}}(\Mat{W}_{1}\Mat{e}_{u}^{(k)} + \Mat{W}_{2}(\Mat{e}_{u}^{(k)}\odot\Mat{e}_{i}^{(k)}))\Big),
\end{aligned}
\end{equation}
where $\textbf{e}_{u}^{(k)}$ and $\textbf{e}_{i}^{(k)}$ respectively denote the refined embedding of user $u$ and item $i$ after $k$ layers propagation, $\sigma$ is the nonlinear activation function,
$\Set{N}_{u}$ denotes the set of items that are interacted by user $u$, $\Set{N}_{i}$ denotes the set of users that interact with item $i$, and $\textbf{W}_1$ and $\textbf{W}_2$ are trainable weight matrix to perform feature transformation in each layer. 
By propagating $L$ layers, NGCF obtains
$L+1$ embeddings to describe a user ($\textbf{e}_u^{(0)}, \textbf{e}_u^{(1)}, ..., \textbf{e}_u^{(L)}$) and an item ($\textbf{e}_i^{(0)}, \textbf{e}_i^{(1)}, ..., \textbf{e}_i^{(L)}$). It then concatenates these $L+1$ embeddings to obtain the final user embedding and item embedding, using inner product to generate the prediction score. 

NGCF largely follows the standard GCN~\cite{GCN}, including the use of nonlinear activation function $\sigma(\cdot)$ and feature transformation matrices $\textbf{W}_1$ and $\textbf{W}_2$. However, we argue that the two operations are not as useful for collaborative filtering. In  semi-supervised node classification, each node has rich semantic features as input, such as the title and abstract words of a paper. Thus performing multiple layers of nonlinear transformation is beneficial to feature learning. Nevertheless, in collaborative filtering, each node of user-item interaction graph only has an ID as input which has no concrete semantics. In this case, performing multiple nonlinear transformations will not contribute to learn better features; even worse, it may add the difficulties to train well. In the next subsection, we provide empirical evidence on this argument. 

\begin{table}[t]
\caption{Performance of NGCF and its three variants.
}
\vspace{-10px}
\label{tab:pre}
\begin{tabular}{l|c c|c c}
\hline
 & \multicolumn{2}{c|}{\textbf{Gowalla}} & \multicolumn{2}{c}{\textbf{Amazon-Book}} \\ \hline
 & \textbf{recall} & \textbf{ndcg} &  \textbf{recall} & \textbf{ndcg} \\ \hline\hline
NGCF &  0.1547 & 0.1307 & 0.0330 & 0.0254 \\ \hline 
NGCF-f    & 0.1686 & 0.1439 & 0.0368 & 0.0283 \\ 
NGCF-n & 0.1536 & 0.1295 & 0.0336 & 0.0258 \\  
NGCF-fn   & 0.1742 & 0.1476 & 0.0399 & 0.0303 \\ \hline
\end{tabular}
\end{table}

\begin{figure*}[t]
	\centering
	\subfigure[Training loss on Gowalla]{\includegraphics[width=0.235\textwidth]{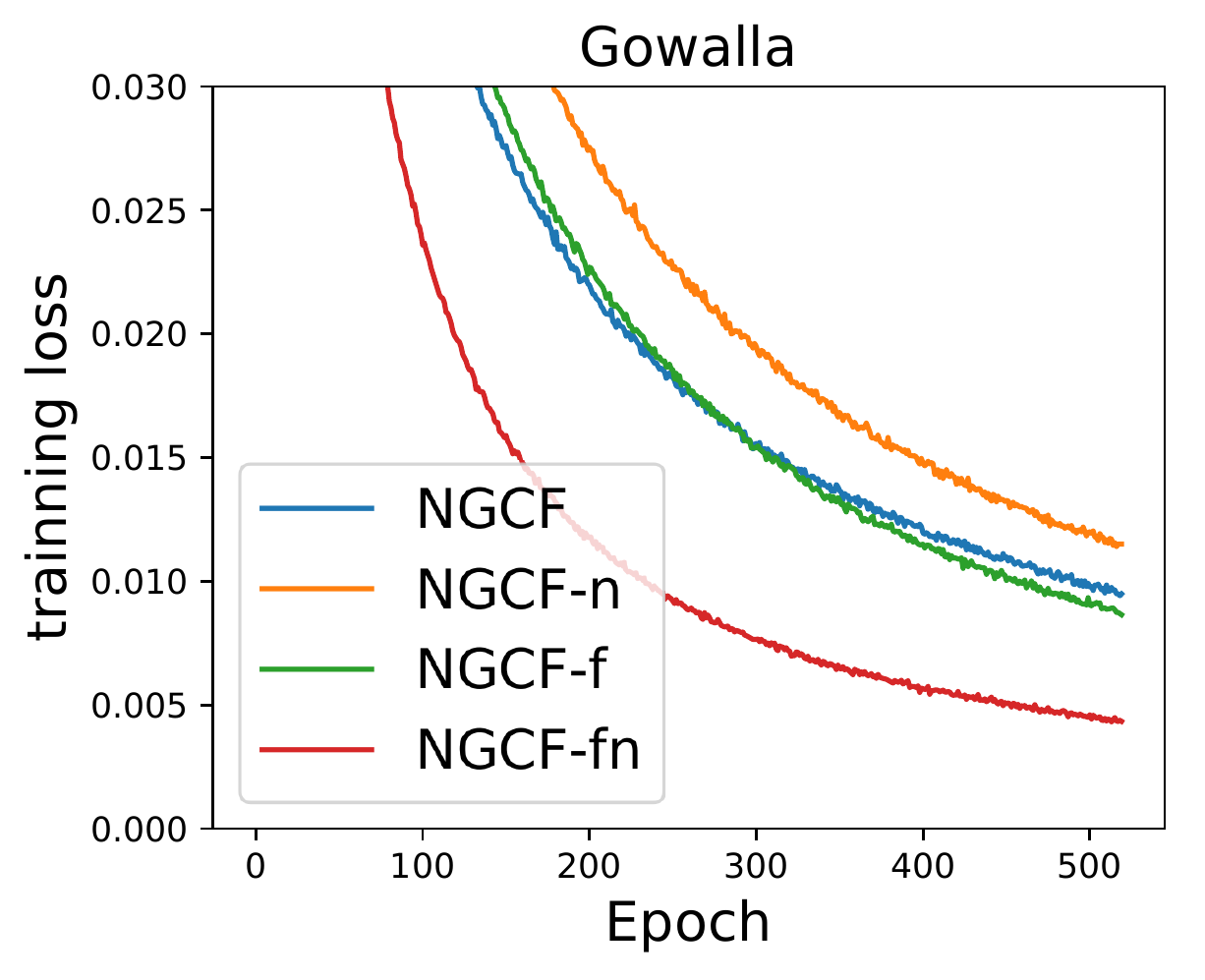}}
	\subfigure[Testing recall on Gowalla]{\includegraphics[width=0.235\textwidth]{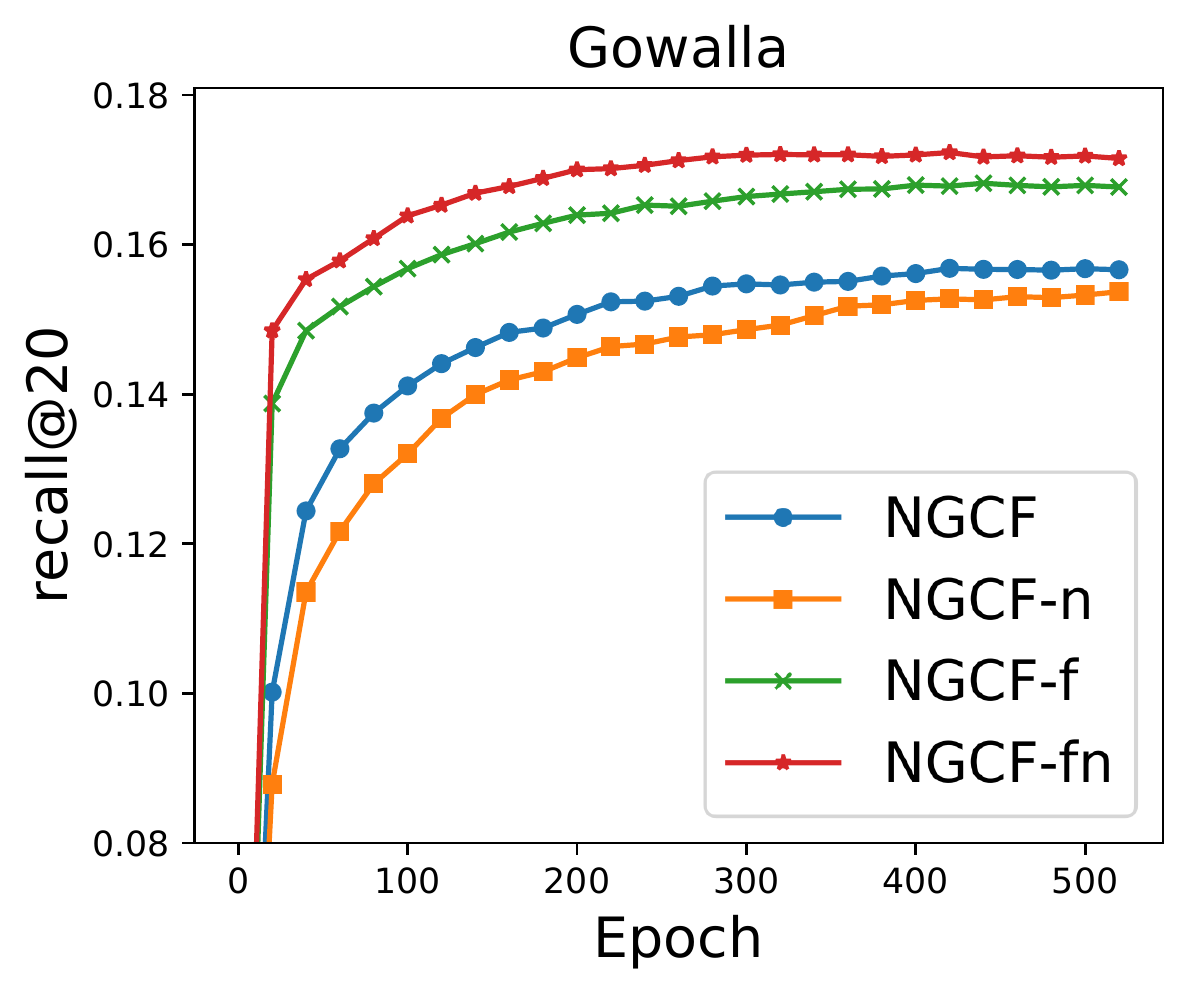}}
	\subfigure[Training loss on Amazon-Book]{\includegraphics[width=0.235\textwidth]{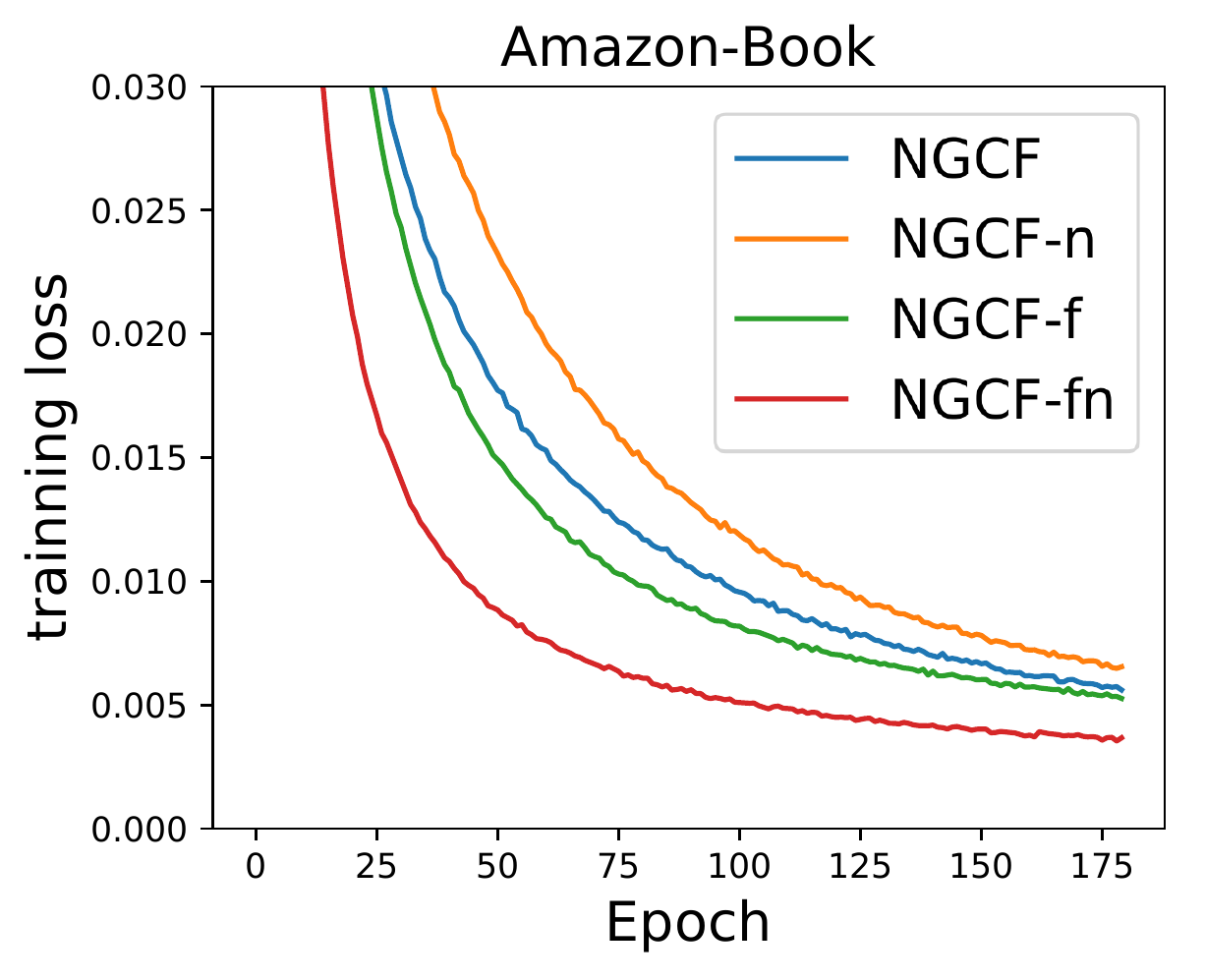}}
	\subfigure[Testing recall on Amazon-Book]{\includegraphics[width=0.235\textwidth]{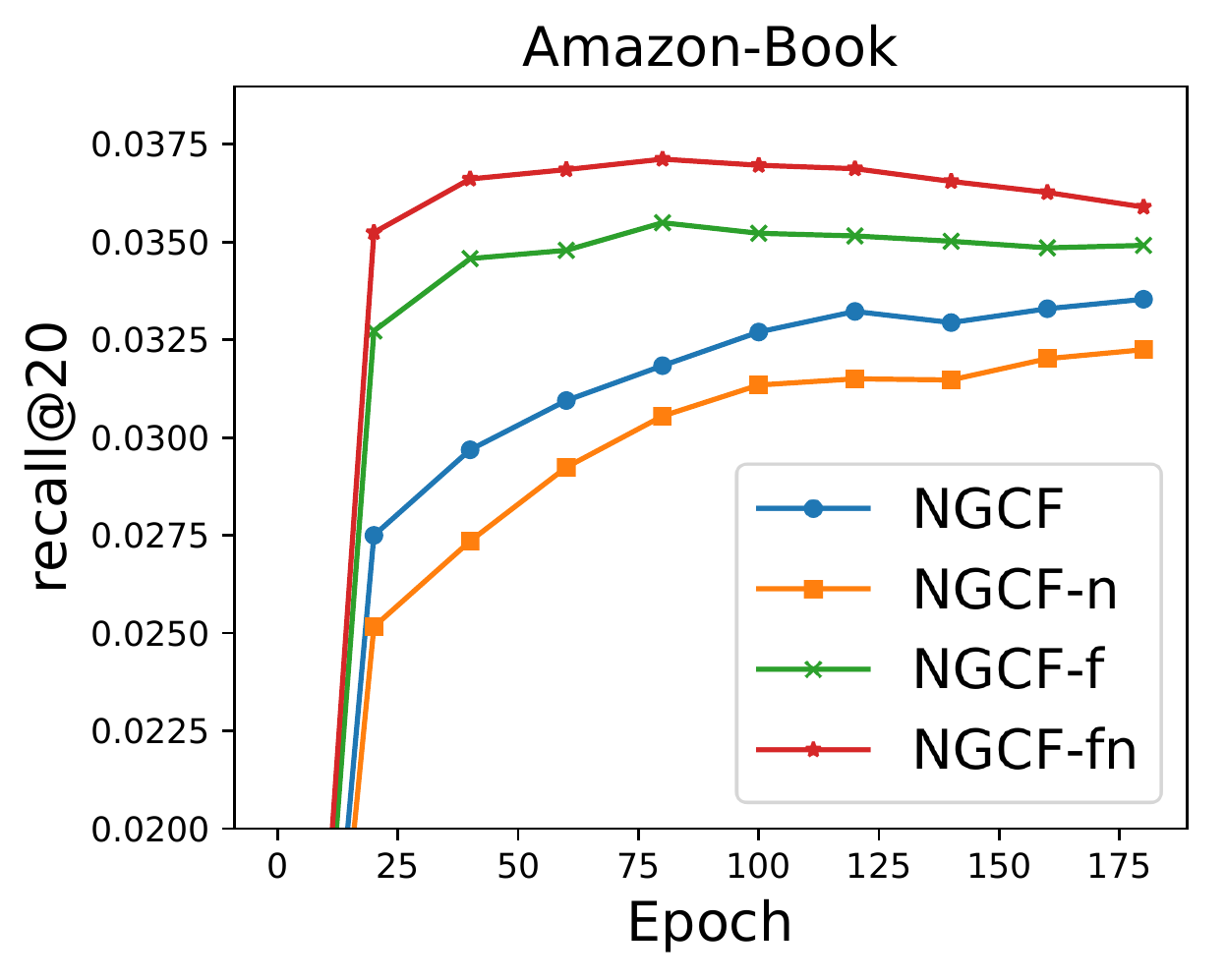}}
	\vspace{-15pt}
	\caption{Training curves (training loss and testing recall) of NGCF and its three simplified variants.} \vspace{-10pt}
	\label{fig:train-epochs}
\end{figure*}

\subsection{Empirical Explorations on NGCF} \label{ss:ngcf_study}
We conduct ablation studies on NGCF to explore the effect of nonlinear activation and feature transformation. We use the codes released by the authors of NGCF\footnote{\url{https://github.com/xiangwang1223/neural_graph_collaborative_filtering}}, running experiments on the same data splits and evaluation protocol to keep the comparison as fair as possible. 
Since the core of GCN is to refine embeddings by propagation, we are more interested in the embedding quality under the same embedding size. Thus, we change the way of obtaining final embedding from concatenation (i.e., $\Mat{e}^{*}_{u}= \Mat{e}^{(0)}_{u}\Vert\cdots\Vert\Mat{e}^{(L)}_{u}$) to sum (i.e., $\Mat{e}^{*}_{u}= \Mat{e}^{(0)}_{u} + \cdots + \Mat{e}^{(L)}_{u}$).
Note that this change has little effect on NGCF's performance, but makes the following ablation studies more indicative of the embedding quality refined by GCN. 

We implement three simplified variants of NGCF:
\begin{itemize}[leftmargin=*]
\item NGCF-f, which removes the feature transformation matrices $\textbf{W}_1$ and $\textbf{W}_2$.
\item NGCF-n, which removes the non-linear activation function $\sigma$.
\item NGCF-fn, which removes both the feature transformation matrices and non-linear activation function. 
\end{itemize}

For the three variants, we keep all hyper-parameters (e.g., learning rate, regularization coefficient, dropout ratio, etc.)  same as the optimal settings of NGCF. 
We report the results of the 2-layer setting on the Gowalla and Amazon-Book datasets in Table \ref{tab:pre}. 
As can be seen, removing feature transformation (i.e., NGCF-f) leads to consistent improvements over NGCF on all three datasets. In contrast, removing nonlinear activation does not affect the accuracy that much. However, if we remove nonlinear activation on the basis of removing feature transformation (i.e., NGCF-fn), the performance is improved significantly. From these observations, we conclude the findings that: 

(1) Adding feature transformation imposes negative effect on NGCF, since removing it in both models of NGCF and NGCF-n improves the performance significantly; 

(2) Adding nonlinear activation affects slightly when feature transformation is included, but it imposes negative effect when feature transformation is disabled. 

(3) As a whole, feature transformation and nonlinear activation impose rather negative effect on NGCF, since by removing them simultaneously, NGCF-fn demonstrates large improvements over NGCF (9.57\% relative improvement on recall). 

To gain more insights into the scores obtained in Table \ref{tab:pre} and understand why NGCF deteriorates with the two operations, we plot the curves of model status recorded by training loss and testing recall in Figure \ref{fig:train-epochs}. 
As can be seen, NGCF-fn achieves a much lower training loss than NGCF, NGCF-f, and NGCF-n along the whole training process. Aligning with the curves of testing recall, we find that such lower training loss successfully transfers to better recommendation accuracy.
The comparison between NGCF and NGCF-f shows the similar trend, except that the improvement margin is smaller. 

From these evidences, we can draw the conclusion that the deterioration of NGCF stems from the training difficulty, rather than overfitting. 
Theoretically speaking, NGCF has higher representation power than NGCF-f, since setting the weight matrix $\textbf{W}_1$ and $\textbf{W}_2$ to identity matrix $\textbf{I}$ can fully recover the NGCF-f model. However, in practice, NGCF demonstrates higher training loss and worse generalization performance than NGCF-f. 
And the incorporation of nonlinear activation further aggravates the discrepancy between representation power and generalization performance. To round out this section, we claim that when designing model for recommendation, it is important to perform rigorous ablation studies to be clear about the impact of each operation. Otherwise, including less useful operations will complicate the model unnecessarily, increase the training difficulty, and even degrade model effectiveness. 
\section{Method}\label{sec:method}
The former section demonstrates that NGCF is a heavy and burdensome GCN model for  collaborative filtering.
Driven by these findings, we set the goal of developing a light yet effective model by including the most essential ingredients of GCN for recommendation. 
The advantages of being simple are several-fold --- more interpretable, practically easy to train and maintain, technically easy to analyze the model behavior and revise it towards more effective directions, and so on. 

\begin{figure}[t]
	\centering
	\small	
	\includegraphics[width=0.48\textwidth]{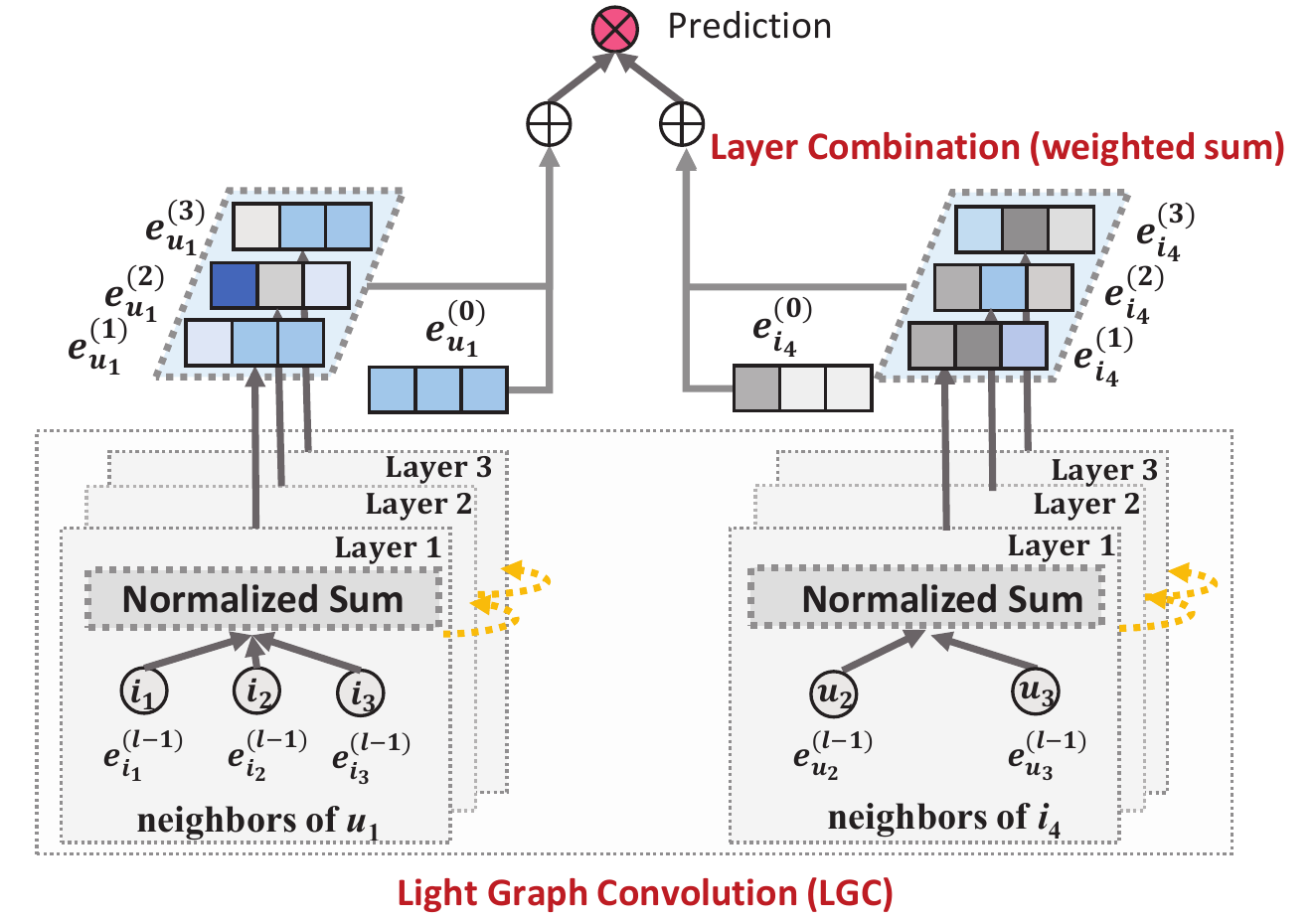}\vspace{-5pt}
	\caption{An illustration of LightGCN model architecture. 
	In LGC, only the normalized sum of neighbor embeddings is performed towards next layer; other operations like self-connection, feature transformation, and nonlinear activation are all removed, which largely simplifies GCNs. In Layer Combination, we sum over the embeddings at each layer to obtain the final representations.}\vspace{-10pt}
	\label{fig:LightGCN}
\end{figure}

In this section, we first present our designed \textit{Light Graph Convolution Network} (LightGCN) model, as illustrated in Figure~\ref{fig:LightGCN}.
We then provide an in-depth analysis of LightGCN to show the rationality behind its simple design. 
Lastly, we describe how to do model training for recommendation. 

\subsection{LightGCN}
The basic idea of GCN is to learning representation for nodes by smoothing features over the graph~\cite{GCN,SGCN}. 
To achieve this, it performs graph convolution iteratively, i.e., aggregating the features of neighbors as the new representation of a target node. 
Such neighborhood aggregation can be abstracted as:
\begin{equation}
    \textbf{e}_{u}^{(k+1)} = \text{AGG} ( \textbf{e}_{u}^{(k)},  \{ \textbf{e}_i^{(k)} : i \in \mathcal{N}_u \}).
\end{equation}
The AGG is an aggregation function --- the core of graph convolution --- that considers the $k$-th layer's representation of the target node and its neighbor nodes. Many work have specified the AGG, such as the weighted sum aggregator in GIN~\cite{GIN}, LSTM aggregator in GraphSAGE~\cite{GraphSAGE}, and bilinear interaction aggregator in BGNN~\cite{BGNN} etc. However, most of the work ties feature transformation or nonlinear activation with the AGG function. Although they perform well on node or graph classification tasks that have semantic input features, they could be burdensome for collaborative filtering (see preliminary results in Section~\ref{ss:ngcf_study}).

\subsubsection{Light Graph Convolution (LGC)}
In LightGCN, we adopt the simple weighted sum aggregator and abandon the use of feature transformation and nonlinear activation. The graph convolution operation (a.k.a., propagation rule~\cite{NGCF}) in LightGCN is defined as:
\begin{equation}\label{eq:LGC}
\begin{aligned}
    \textbf{e}_{u}^{(k+1)} &= \sum_{i \in \mathcal{N}_u} \frac{1}{\sqrt{|\mathcal{N}_u|}\sqrt{|\mathcal{N}_i|}} \textbf{e}_i^{(k)},  \\
    \textbf{e}_{i}^{(k+1)} &= \sum_{u \in \mathcal{N}_i} \frac{1}{\sqrt{|\mathcal{N}_i|}\sqrt{|\mathcal{N}_u|}} \textbf{e}_u^{(k)}.
\end{aligned}
\end{equation}
The symmetric normalization term $\frac{1}{\sqrt{|\mathcal{N}_u|}\sqrt{|\mathcal{N}_i|}}$ follows the design of standard GCN~\cite{GCN}, which can avoid the scale of embeddings increasing with graph convolution operations; other choices can also be applied here, such as the $L_1$ norm, while empirically we find this symmetric normalization has good performance (see experiment results in Section~\ref{ss:ablation-symmetric}). 

It is worth noting that in LGC, we aggregate only the connected neighbors and do not integrate the target node itself (i.e., self-connection). This is different from most existing graph convolution operations~\cite{NGCF,GCN,GAT,GraphSAGE, BGNN} that typically aggregate extended neighbors and need to handle the self-connection specially. 
The layer combination operation, to be introduced in the next subsection, essentially captures the same effect as self-connections. Thus, there is no need in LGC to include self-connections. 

\subsubsection{Layer Combination and Model Prediction} In LightGCN, the only trainable model parameters are the embeddings at the 0-th layer, i.e., $\textbf{e}_{u}^{(0)}$ for all users and  $\textbf{e}_{i}^{(0)}$ for all items. When they are given, the embeddings at higher layers can be computed via LGC defined in Equation~(\ref{eq:LGC}). 
After $K$ layers LGC, we further combine the embeddings obtained at each layer to form the final representation of a user (an item):
\begin{equation}
\begin{aligned}
    \textbf{e}_{u} = \sum_{k=0}^K \alpha_k \textbf{e}_u^{(k)}; \quad \textbf{e}_{i} = \sum_{k=0}^K \alpha_k \textbf{e}_i^{(k)},
\end{aligned}
\end{equation}
where $\alpha_k \geq 0$ denotes the importance of the $k$-th layer embedding in constituting the final embedding. 
It can be treated as a hyper-parameter to be tuned manually, or as a model parameter (e.g., output of an attention network~\cite{ACF}) to be optimized automatically. In our experiments, we find that setting $\alpha_k$ uniformly as $1 / (K+1)$ leads to good performance in general. Thus we do not design special component to optimize $\alpha_k$, to avoid complicating LightGCN unnecessarily and to keep its simplicity.  
The reasons that we perform layer combination to get final representations are three-fold. (1) With the increasing of the number of layers, the embeddings will be over-smoothed~\cite{DeepInsights}. Thus simply using the last layer is problematic. (2) The embeddings at different layers capture different semantics. E.g., the first layer enforces smoothness on users and items that have interactions, the second layer smooths users (items) that have overlap on interacted items (users), and higher-layers capture higher-order proximity~\cite{NGCF}. Thus combining them will make the representation more comprehensive. (3) Combining embeddings at different layers with weighted sum captures the effect of graph convolution with self-connections, an important trick in GCNs (proof sees Section~\ref{ss:relation-sgcn}).

The model prediction is defined as the inner product of user and item final representations: 
\begin{equation}
    \hat{y}_{ui} = \textbf{e}_u ^T \textbf{e}_i,
\end{equation}
which is used as the ranking score for recommendation generation. 

\subsubsection{Matrix Form} We provide the matrix form of LightGCN to facilitate implementation and discussion with existing models. Let the user-item interaction matrix be $\textbf{R} \in \mathbb{R}^{M\times N}$ where $M$ and $N$ denote the number of users and items, respectively, and each entry $R_{ui}$ is 1 if $u$ has interacted with item $i$ otherwise 0. We then obtain the adjacency matrix of the user-item graph as 
\begin{equation}
    \textbf{A}=
    \left(\begin{matrix}
    \textbf{0} & \textbf{R}\\
    \textbf{R}^{T} & \textbf{0}\\
    \end{matrix} \right),
\end{equation}
Let the $0$-th layer embedding matrix be $\textbf{E}^{(0)} \in \mathbb{R}^{ (M+N) \times T} $, where $T$ is the embedding size.  
Then we can obtain the matrix equivalent form of LGC as:
\begin{equation}
    \textbf{E}^{(k+1)} = (\textbf{D}^{-\frac{1}{2}} \textbf{A} \textbf{D}^{-\frac{1}{2}}) \textbf{E}^{(k)},
\end{equation}
where $\textbf{D}$ is a $(M+N)\times (M+N)$ diagonal matrix, in which each entry $D_{ii}$ denotes the number of nonzero entries in the $i$-th row vector of the  adjacency matrix $\textbf{A}$ (also named as degree matrix). Lastly, we get the final embedding matrix used for model prediction as:
\begin{equation}\label{eq:E_LGCN}
\begin{aligned}
    \textbf{E} &= \alpha_0 \textbf{E}^{(0)} + \alpha_1 \textbf{E}^{(1)} + \alpha_2 \textbf{E}^{(2)} + ... + \alpha_K \textbf{E}^{(K)} \\
    &= \alpha_0 \textbf{E}^{(0)} + \alpha_1 \Tilde{\textbf{A}}\textbf{E}^{(0)} + \alpha_2 \Tilde{\textbf{A}}^2\textbf{E}^{(0)} + ... + \alpha_K \Tilde{\textbf{A}}^K\textbf{E}^{(0)},
\end{aligned}
\end{equation}
where $\Tilde{\textbf{A}} = \textbf{D}^{-\frac{1}{2}} \textbf{A} \textbf{D}^{-\frac{1}{2}}$ is the symmetrically normalized matrix.

\subsection{Model Analysis}
We conduct model analysis to demonstrate the rationality behind the simple design of LightGCN. 
First we discuss the connection with the Simplified GCN (SGCN)~\cite{SGCN}, which is a recent linear GCN model that integrates self-connection into graph convolution; this analysis shows that by doing layer combination, LightGCN subsumes the effect of self-connection thus there is no need for LightGCN to add self-connection in adjacency matrix. 
Then we discuss the relation with the Approximate Personalized Propagation of Neural Predictions (APPNP)~\cite{ICLR19-APPNP}, which is recent GCN variant that addresses oversmoothing by inspiring from Personalized PageRank~\cite{haveliwala2002topic}; this analysis shows the underlying equivalence between LightGCN and APPNP, thus our LightGCN enjoys the sames benefits in propagating long-range with controllable oversmoothing. 
Lastly we analyze the second-layer LGC to show how it smooths a user with her second-order neighbors, providing more insights into the working mechanism of LightGCN.  

\subsubsection{Relation with SGCN}\label{ss:relation-sgcn} In \cite{SGCN}, the authors argue the unnecessary complexity of GCN for node classfication and propose SGCN, which simplifies GCN by removing nonlinearities and collapsing the weight matrices to one weight matrix. The graph convolution in SGCN is defined as\footnote{The weight matrix in SGCN can be absorbed into the 0-th layer embedding parameters, thus it is omitted in the analysis.}:
\begin{equation}
    \textbf{E}^{(k+1)} = (\textbf{D}+\textbf{I})^{-\frac{1}{2}} (\textbf{A}+\textbf{I}) (\textbf{D}+\textbf{I})^{-\frac{1}{2}} \textbf{E}^{(k)},
\end{equation}
where $\textbf{I}\in \mathbb{R}^{(M+N)\times (M+N)}$ is an identity matrix, which is added on $\textbf{A}$ to include self-connections. 
In the following analysis, we omit the $(\textbf{D}+\textbf{I})^{-\frac{1}{2}}$ terms for simplicity, since they only re-scale embeddings. In SGCN, the embeddings obtained at the last layer are used for downstream prediction task, which can be expressed as:
\begin{equation}\label{eq:SGCN}
\begin{aligned}
    \textbf{E}^{(K)} &= (\textbf{A}+\textbf{I}) \textbf{E}^{(K-1)} = (\textbf{A}+\textbf{I})^K \textbf{E}^{(0)} \\
    &= \binom{K}{0}\textbf{E}^{(0)} + \binom{K}{1} \textbf{A} \textbf{E}^{(0)} + \binom{K}{2} \textbf{A}^2 \textbf{E}^{(0)} + ... + \binom{K}{K} \textbf{A}^K \textbf{E}^{(0)}.
\end{aligned}
\end{equation}
The above derivation shows that, inserting self-connection into $\textbf{A}$ and propagating embeddings on it, is essentially equivalent to a weighted sum of the embeddings propagated at each LGC layer. 

\subsubsection{Relation with APPNP}\label{ss:APPNP} In a recent work \cite{ICLR19-APPNP}, the authors
connect GCN with Personalized PageRank~\cite{haveliwala2002topic}, inspiring from which they propose a GCN variant named APPNP that can propagate long range without the risk of oversmoothing. 
Inspired by the teleport design in Personalized PageRank, APPNP complements each propagation layer with the starting features (i.e., the 0-th layer embeddings), which can balance the need of preserving locality (i.e., staying close to the root node to alleviate oversmoothing) and leveraging the information from a large neighborhood. The propagation layer in APPNP is defined as:
\begin{equation}
    \textbf{E}^{(k+1)} = \beta \textbf{E}^{(0)} + (1-\beta) \Tilde{\textbf{A}} \textbf{E}^{(k)},
\end{equation}
where $\beta$ is the teleport probability to control the retaining of starting features in the propagation, and $\Tilde{\textbf{A}}$ denotes the normalized adjacency matrix. In APPNP, the last layer is used for final prediction, i.e., 
\begin{equation}
\begin{aligned}
    \textbf{E}^{(K)} &= \beta \textbf{E}^{(0)} + (1-\beta) \Tilde{\textbf{A}} \textbf{E}^{(K-1)}, \\
    &= \beta \textbf{E}^{(0)} + \beta (1-\beta) \Tilde{\textbf{A}} \textbf{E}^{(0)} + (1-\beta)^2 \Tilde{\textbf{A}}^2 \textbf{E}^{(K-2)} \\
    &= \beta \textbf{E}^{(0)} + \beta (1-\beta) \Tilde{\textbf{A}} \textbf{E}^{(0)} + \beta(1-\beta)^2 \Tilde{\textbf{A}}^2 \textbf{E}^{(0)} + ... + (1-\beta)^K \Tilde{\textbf{A}}^K \textbf{E}^{(0)}.
\end{aligned}
\end{equation}
Aligning with Equation~(\ref{eq:E_LGCN}), we can see that by setting $\alpha_k$ accordingly, LightGCN can fully recover the prediction embedding used by APPNP. 
As such, LightGCN shares the strength of APPNP in combating oversmoothing --- by setting the $\alpha$ properly, we allow using a large $K$ for long-range modeling with controllable oversmoothing. 

Another minor difference is that APPNP adds self-connection into the adjacency matrix. However, as we have shown before, this is redundant due to the weighted sum of different layers. 

\subsubsection{Second-Order Embedding Smoothness} \label{ss:second-order}
Owing to the linearity and simplicity of LightGCN, we can draw more insights into how does it smooth embeddings. Here we analyze a 2-layer LightGCN to demonstrate its rationality. Taking the user side as an example, intuitively, the second layer smooths users that have overlap on the interacted items. More concretely, we have:
\begin{equation}
\begin{aligned}
    \textbf{e}_{u}^{(2)} &= \sum_{i \in \mathcal{N}_u} \frac{1}{\sqrt{|\mathcal{N}_u|}\sqrt{|\mathcal{N}_i|}} \textbf{e}_i^{(1)} = \sum_{i \in \mathcal{N}_u} \frac{1}{|\mathcal{N}_i|} \sum_{v\in \mathcal{N}_i} \frac{1}{\sqrt{|\mathcal{N}_u|}\sqrt{|\mathcal{N}_v|} } \textbf{e}_{v}^{(0)}. \\ 
\end{aligned}
\end{equation}
We can see that, if another user $v$ has co-interacted with the target user $u$, the smoothness strength of $v$ on $u$ is measured by the coefficient (otherwise 0):
\begin{equation}\label{eq:c_vu}
    c_{v -> u} = \frac{1}{\sqrt{|\mathcal{N}_u|}\sqrt{|\mathcal{N}_v|}} \sum_{i \in \mathcal{N}_u \cap \mathcal{N}_v} \frac{1}{|\mathcal{N}_i|}. 
\end{equation}
This coefficient is rather interpretable: the influence of a second-order neighbor $v$ on $u$ is determined by 1) the number of co-interacted items, the more the larger; 2) the popularity of the co-interacted items, the less popularity (i.e., more indicative of user personalized preference) the larger; and 3) the activity of $v$, the less active the larger. 
Such interpretability well caters for the assumption of CF in measuring user similarity~\cite{CSE,Wang:2006} and evidences the reasonability of LightGCN. 
Due to the symmetric formulation of LightGCN, we can get similar analysis on the item side. 

\subsection{Model Training}
The trainable parameters of LightGCN are only the embeddings of the 0-th layer, i.e.,  $\Theta=\{\textbf{E}^{(0)}\}$; in other words, the model complexity is same as the standard matrix factorization (MF).
We employ the \textit{Bayesian Personalized Ranking} (BPR) loss~\cite{BPRMF}, which is a pairwise loss that encourages the prediction of an observed entry to be higher than its unobserved counterparts:
\begin{equation}
 L_{BPR} = - \sum_{u=1}^{M} \sum_{i \in \mathcal{N}_u} \sum_{j \notin \mathcal{N}_u} \ln \sigma (\hat{y}_{ui} - \hat{y}_{uj} ) + \lambda ||\textbf{E}^{(0)}||^2
\end{equation}
where $\lambda$ controls the $L_2$ regularization strength. We employ the Adam~\cite{Adam} optimizer and use it in a mini-batch manner.
We are aware of other advanced negative sampling strategies which might improve the LightGCN training, such as the hard negative sampling~\cite{rendle2014improving} and adversarial sampling~\cite{Ding2019IJCAI}. We leave this extension in the future since it is not the focus of this work. 

Note that we do not introduce dropout mechanisms, which are commonly used in GCNs and NGCF. The reason is that we do not have feature transformation weight matrices in LightGCN, thus enforcing $L_2$ regularization on the embedding layer is sufficient to prevent overfitting. 
This showcases LightGCN's advantages of being simple --- it is easier to train and tune than NGCF which additionally requires to tune two dropout ratios (node dropout and message dropout) and normalize the embedding of each layer to unit length. 

Moreover, it is technically viable to also learn the layer combination coefficients $\{\alpha_k\}_{k=0}^K$, or parameterize them with an attention network. However, we find that learning $\alpha$ on training data does not lead improvement. This is probably because the training data does not contain sufficient signal to learn good $\alpha$ that can generalize to unknown data. We have also tried to learn $\alpha$ from validation data, as inspired by \cite{lambdaOpt} that learns hyper-parameters on validation data. The performance is slightly improved (less than $1\%$). We leave the exploration of optimal settings of $\alpha$ (e.g., personalizing it for different users and items) as future work. 
\section{Experiments}\label{sec:experiments}
We first describe experimental settings, and then conduct detailed comparison with NGCF~\cite{NGCF}, the method that is most relevant with LightGCN but more complicated (Section~\ref{ss:exp-ngcf}). We next compare with other state-of-the-art methods in Section \ref{ss:exp-SOTA}. To justify the designs in LightGCN and reveal the reasons of its effectiveness, we perform ablation studies and embedding analyses in Section \ref{ss:exp-ablation}. The hyper-parameter study is finally presented in Section \ref{ss:exp-hyper}. 

\begin{table}[t]
\caption{Statistics of the experimented data.}
\vspace{-10px}
\label{tab:dataset}
\begin{tabular}{l|r|r|r|r}
\hline
\multicolumn{1}{c|}{\textbf{Dataset}} & \multicolumn{1}{c|}{\textbf{User \#}} & \multicolumn{1}{c|}{\textbf{Item \#}} & \multicolumn{1}{c|}{\textbf{Interaction \#}} & \multicolumn{1}{c}{\textbf{Density}} \\ \hline\hline
Gowalla & $29,858$ & $40,981$ & $1,027,370$ & $0.00084$ \\ \hline
Yelp2018 & $31,668$ & $38,048$ & $1,561,406$ & $0.00130$ \\ \hline
Amazon-Book & $52,643$ & $91,599$ & $2,984,108$ & $0.00062$ \\ \hline
\end{tabular}
\vspace{-15px}
\end{table}

\subsection{Experimental Settings}\label{ss:setting}
To reduce the experiment workload and keep the comparison fair, we closely follow the settings of the NGCF work~\cite{NGCF}. We request the experimented datasets (including train/test splits) from the authors, for which the statistics are shown in Table~\ref{tab:dataset}. 
The Gowalla and Amazon-Book are exactly the same as the NGCF paper used, so we directly use the results in the NGCF paper. The only exception is the Yelp2018 data, which is a revised version. According to the authors, the previous version did not filter out cold-start items in the testing set, and they shared us the revised version only. Thus we re-run NGCF on the Yelp2018 data.
The evaluation metrics are recall@20 and ndcg@20 computed by the all-ranking protocol --- all items that are not interacted by a user are the candidates. 

\begin{table*}[t]
\caption{Performance comparison between NGCF and LightGCN at different layers.}
\vspace{-10px}
\label{tab:overall}
\resizebox{0.9\textwidth}{!}{
\begin{tabular}{c|c|l | l|l | l| l | l}
\hline
 \multicolumn{2}{c|}{\textbf{Dataset}}
 & \multicolumn{2}{c|}{\textbf{Gowalla}} & \multicolumn{2}{c|}{\textbf{Yelp2018}} & \multicolumn{2}{c}{\textbf{Amazon-Book}} \\ \hline
 \textbf{Layer \#} & \textbf{Method} 
 & \textbf{recall} & \textbf{ndcg} & \textbf{recall} & \textbf{ndcg} & \textbf{recall} & \textbf{ndcg} \\ \hline\hline
\multirow{2}{*}{\textbf{1 Layer}} & NGCF &  
0.1556 & 0.1315 & 0.0543 & 0.0442 & 0.0313 & 0.0241
 \\ 
&LightGCN & 0.1755(+12.79\%) & 0.1492(+13.46\%) & 0.0631(+16.20\%) & 0.0515(+16.51\%) & 0.0384(+22.68\%) & 0.0298(+23.65\%) \\ \hline
\multirow{2}{*}{\textbf{2 Layers}} & NGCF &  
0.1547 & 0.1307 & 0.0566 & 0.0465 & 0.0330 &	0.0254 \\
&LightGCN & 0.1777(+14.84\%) & 0.1524(+16.60\%) & 0.0622(+9.89\%) & 0.0504(+8.38\%) & 0.0411(+24.54\%) & 0.0315(+24.02\%) \\  \hline
\multirow{2}{*}{\textbf{3 Layers}} & NGCF &  
0.1569 & 0.1327 & 0.0579 & 0.0477 & 0.0337 & 0.0261 \\ 
&LightGCN & 0.1823(+16.19\%) & 0.1555(+17.18\%) & 0.0639(+10.38\%) & 0.0525(+10.06\%) & 0.0410(+21.66\%) & 0.0318(+21.84\%) \\ \hline
\multirow{2}{*}{\textbf{4 Layers}} & NGCF &  
0.1570 & 0.1327 & 0.0566 & 0.0461 & 0.0344 & 0.0263\\
&LightGCN & 0.1830(+16.56\%) & 0.1550(+16.80\%) & 0.0649(+14.58\%) & 0.0530(+15.02\%) & 0.0406(+17.92\%) & 0.0313(+18.92\%) \\ \hline
\end{tabular}}\vspace{+2pt}
\\\small{*The scores of NGCF on Gowalla and Amazon-Book are directly copied from Table 3 of the NGCF paper (\url{https://arxiv.org/abs/1905.08108})} \vspace{-5pt}
\end{table*}

\subsubsection{Compared Methods} The main competing method is NGCF, which has shown to outperform several methods including GCN-based models GC-MC~\cite{GC-MC} and PinSage~\cite{PinSage}, neural network-based models NeuMF~\cite{NCF} and CMN~\cite{CMN}, and factorization-based models MF~\cite{BPRMF} and HOP-Rec~\cite{HOP-rec}.
As the comparison is done on the same datasets under the same evaluation protocol, we do not further compare with these methods. In addition to NGCF, we further compare with two relevant and competitive CF methods:
\begin{itemize}[leftmargin=*]
\item Mult-VAE~\cite{VACF}. This is an item-based CF method based on the variational autoencoder (VAE). 
It assumes the data is generated from a multinomial distribution and using variational inference for parameter estimation. 
We run the codes released by the authors\footnote{\url{https://github.com/dawenl/vae_cf}}, tuning the dropout ratio in $[0, 0.2, 0.5]$, and the $\beta$ in $[0.2,0.4,0.6,0.8]$. The model architecture is the suggested one in the paper: $600\rightarrow 200 \rightarrow 600$. 
\item GRMF~\cite{rao2015collaborative}. This method smooths matrix factorization by adding the graph Laplacian regularizer.
For fair comparison on item recommendation, we change the rating prediction loss to BPR loss. The objective function of GRMF is:
\begin{equation}
\begin{aligned}
    L &= - \sum_{u=1}^{M} \sum_{i \in \mathcal{N}_u} \Big(\sum_{j \notin \mathcal{N}_u} \ln \sigma (\textbf{e}_u^T \textbf{e}_i - \textbf{e}_u^T \textbf{e}_j ) + \lambda_{g} ||\textbf{e}_u - \textbf{e}_i||^2 \Big)+ \lambda ||\textbf{E}||^2,
\end{aligned}
\end{equation}
where $\lambda_g$ is searched in the range of $[1e^{-5}, 1e^{-4}, ..., 1e^{-1}]$. 
Moreover, we compare with a variant that adds normalization to graph Laplacian: $\lambda_{g} ||\frac{\textbf{e}_u}{\sqrt{|\mathcal{N}_u|}} - \frac{\textbf{e}_i}{\sqrt{|\mathcal{N}_i|}}||^2$, which is termed as GRMF-norm. 
Other hyper-parameter settings are same as LightGCN. 
The two GRMF methods benchmark the performance of smoothing embeddings via Laplacian regularizer, while our LightGCN achieves embedding smoothing in the predictive model. 
\end{itemize}

\subsubsection{Hyper-parameter Settings} 
Same as NGCF, the embedding size is fixed to 64 for all models and the embedding parameters are initialized with the Xavier method~\cite{Xarvier}. We optimize LightGCN with Adam~\cite{Adam} and use the default learning rate of 0.001 and default mini-batch size of 1024 (on Amazon-Book, we increase the mini-batch size to 2048 for speed). The $L_2$ regularization coefficient $\lambda$ is searched in the range of $\{1e^{-6}, 1e^{-5}, ... , 1e^{-2}\}$, and in most cases the optimal value is $1e^{-4}$. 
The layer combination coefficient $\alpha_k$ is uniformly set to $\frac{1}{1+K}$ where $K$ is the number of layers. We test $K$ in the range of 1 to 4, and satisfactory performance can be achieved when $K$ equals to 3. 
The early stopping and validation strategies are the same as NGCF. Typically, 1000 epochs are sufficient for LightGCN to converge. Our implementations are available in both TensorFlow\footnote{\url{https://github.com/kuandeng/LightGCN}} and PyTorch\footnote{\url{https://github.com/gusye1234/pytorch-light-gcn}}.

\begin{figure*}[t]
	\centering
	\subfigure{\includegraphics[width=0.24\textwidth]{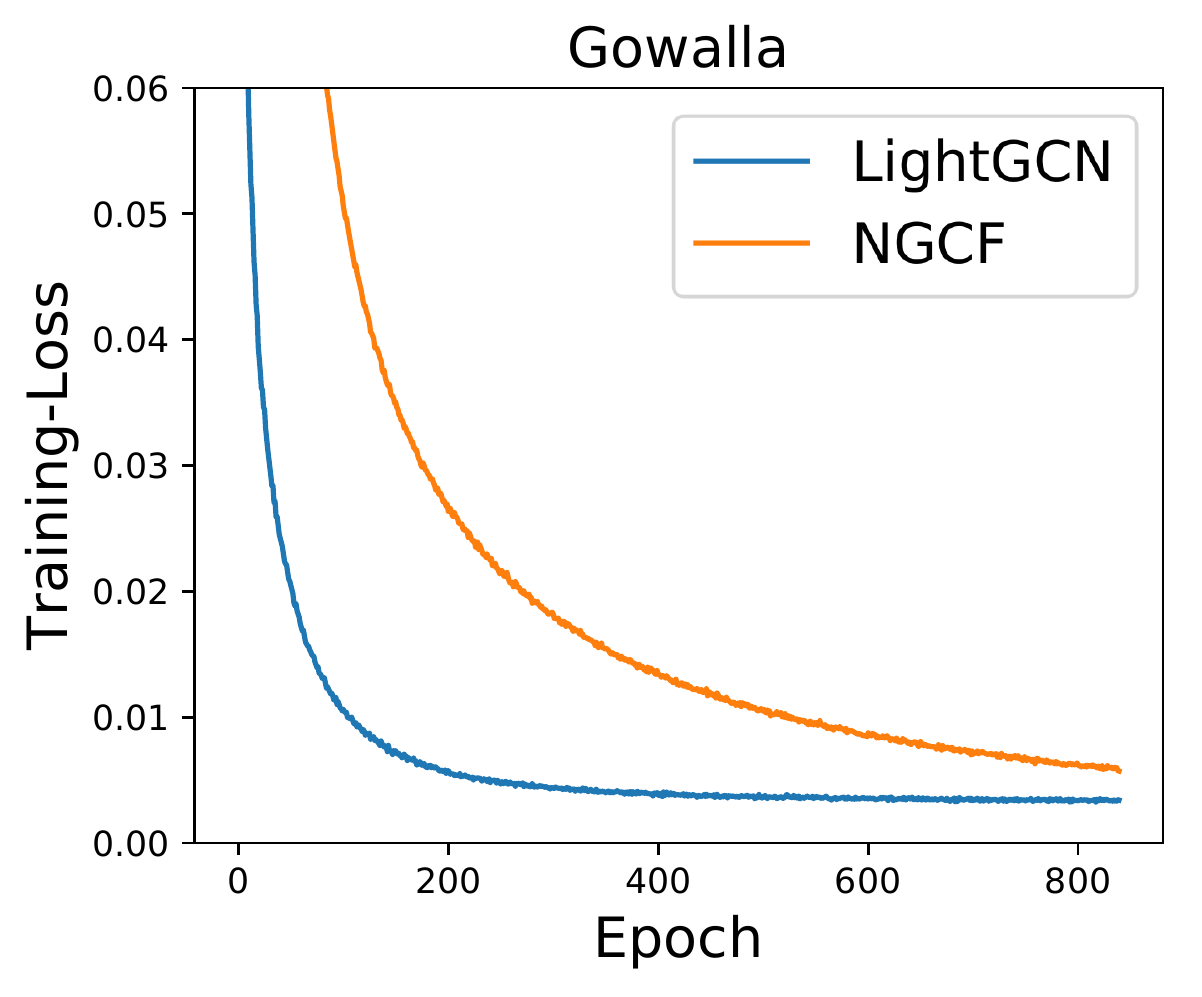}} 
	\subfigure{\includegraphics[width=0.24\textwidth]{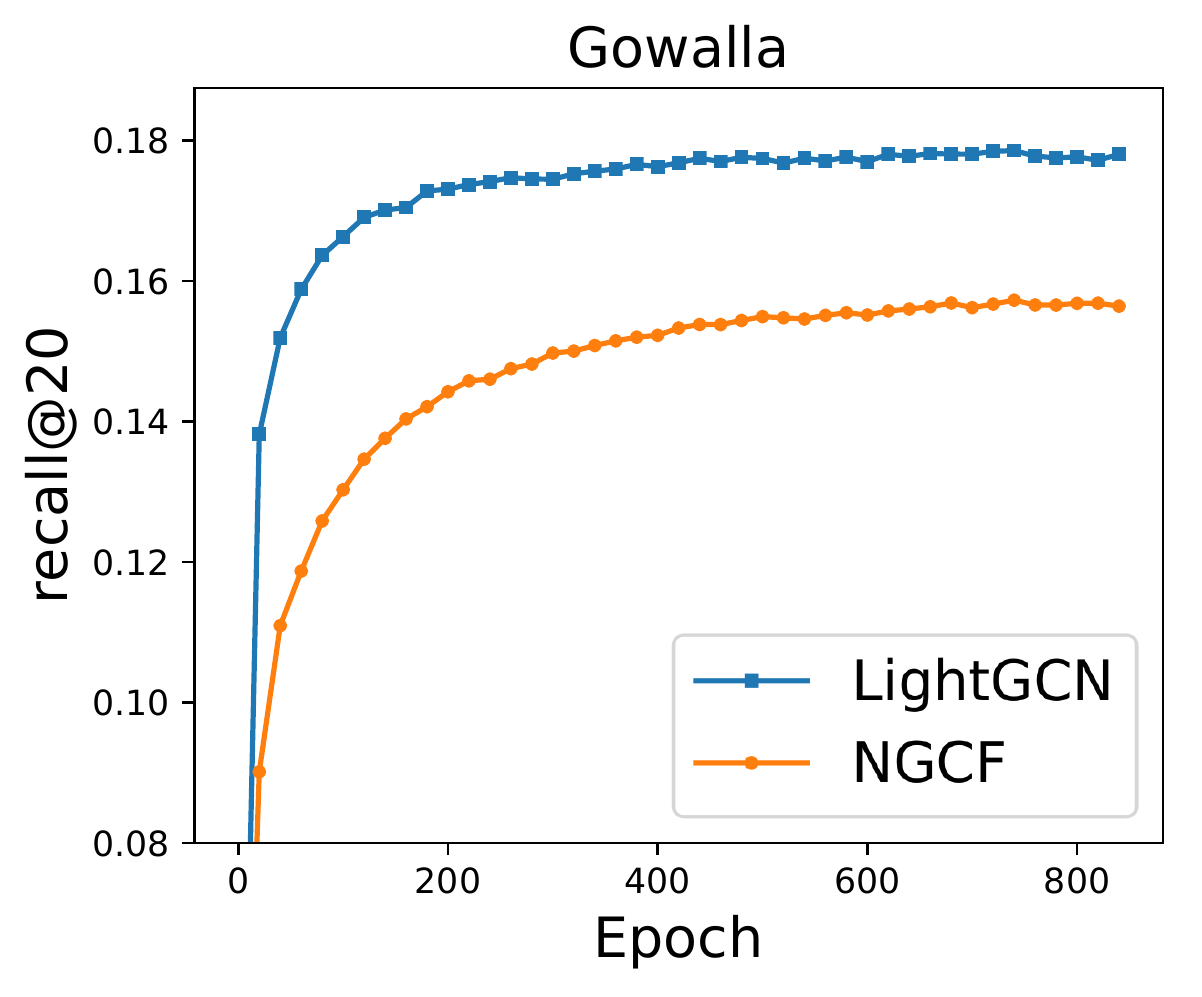}} 
	\subfigure{\includegraphics[width=0.24\textwidth]{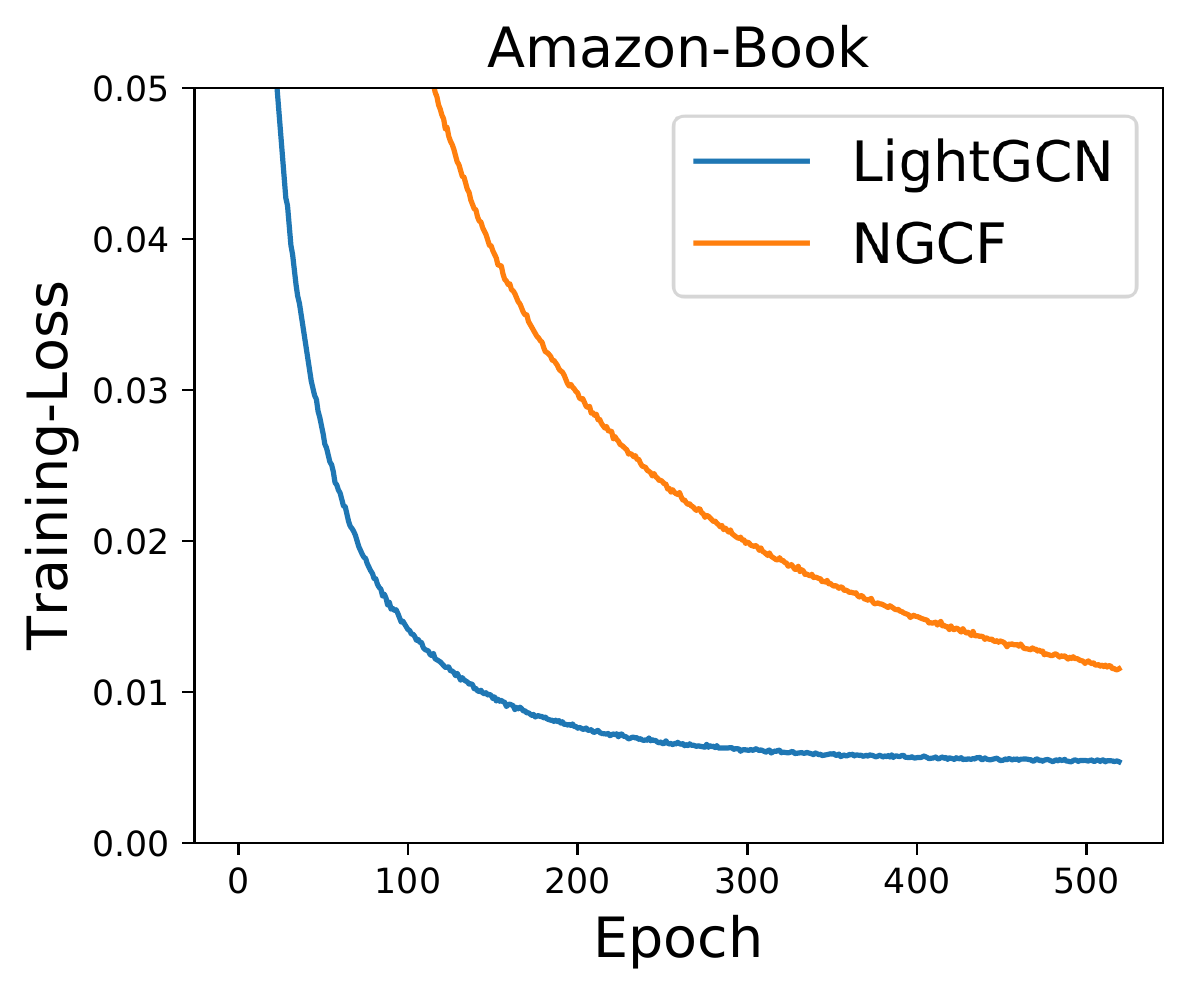}}
	\subfigure{\includegraphics[width=0.24\textwidth]{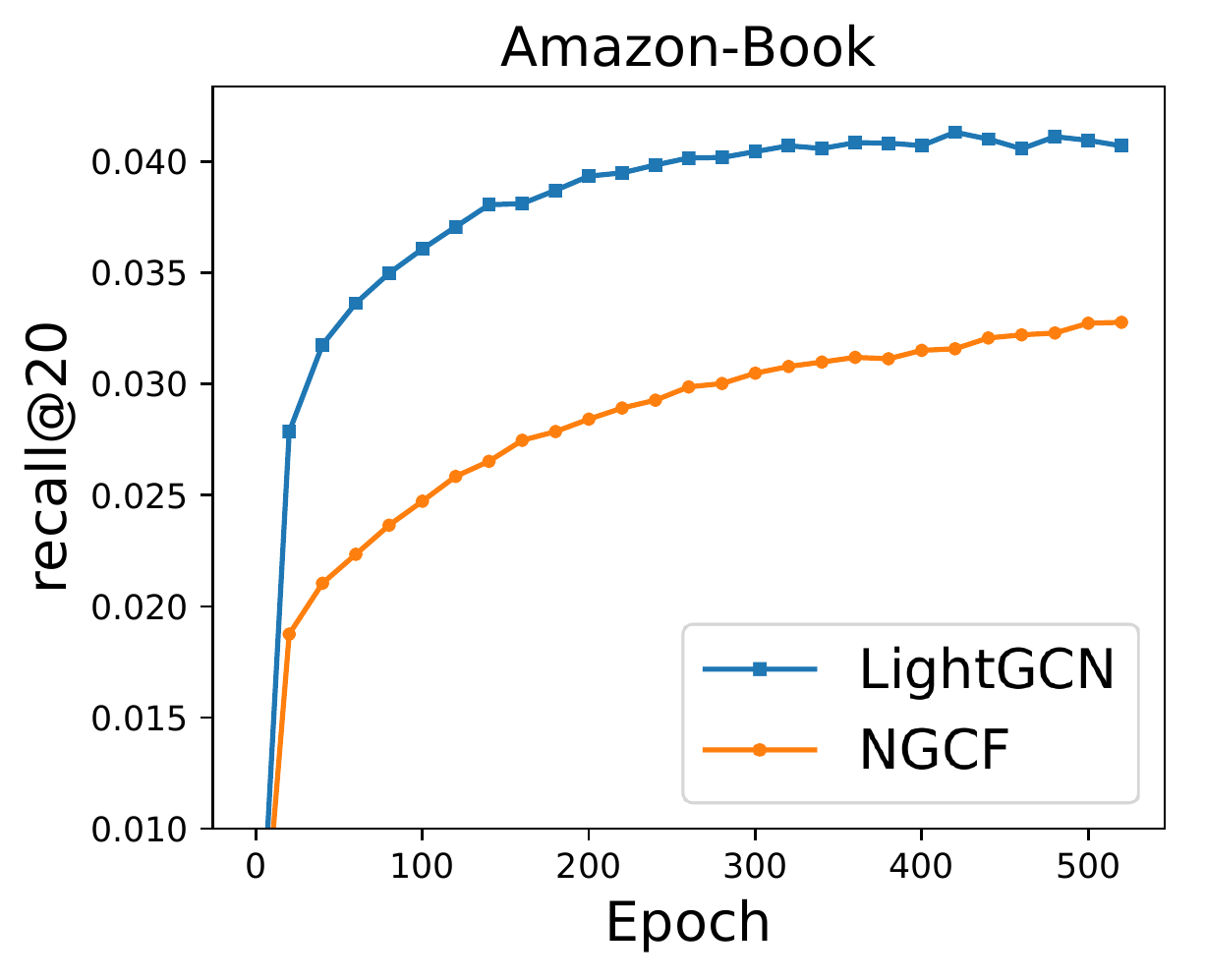}}
	\vspace{-15pt}
	\caption{Training curves of LightGCN and NGCF, which are evaluated by training loss and testing recall per 20 epochs on Gowalla and Amazon-Book (results on Yelp2018 show exactly the same trend which are omitted for space).} \vspace{-10pt}
	\label{fig:curve}
\end{figure*}

\subsection{Performance Comparison with NGCF}\label{ss:exp-ngcf}
We perform detailed comparison with NGCF, recording the performance at different layers (1 to 4) in Table \ref{tab:overall}, which also shows the percentage of relative improvement on each metric. 
We further plot the training curves of training loss and testing recall in Figure~\ref{fig:curve} to reveal the advantages of LightGCN and to be clear of the training process. 
The main observations are as follows:
\begin{itemize}[leftmargin=*]
    \item In all cases, LightGCN outperforms NGCF by a large margin. For example, on Gowalla the highest recall reported in the NGCF paper is 0.1570, while our LightGCN can reach 0.1830 under the 4-layer setting, which is $16.56\%$ higher. On average, the recall improvement on the three datasets is $16.52\%$ and the ndcg improvement is $16.87\%$, which are rather significant.
    \item Aligning Table \ref{tab:overall} with Table \ref{tab:pre} in Section~\ref{sec:preliminaries}, we can see that LightGCN performs better than NGCF-fn, the variant of NGCF that removes feature transformation and nonlinear activation. As NGCF-fn still contains more operations than LightGCN (e.g., self-connection, the interaction between user embedding and item embedding in graph convolution, and dropout), this suggests that these operations might also be useless for NGCF-fn. 
    \item Increasing the number of layers can improve the performance, but the benefits diminish. The general observation is that increasing the layer number from 0 (i.e., the matrix factorization model, results see \cite{NGCF}) to 1 leads to the largest performance gain, and using a layer number of 3 leads to satisfactory performance in most cases. This observation is consistent with NGCF's finding.
    \item Along the training process, LightGCN consistently obtains lower training loss, which indicates that LightGCN fits the training data better than NGCF. Moreover, the lower training loss successfully transfers to better testing accuracy, indicating the strong generalization power of LightGCN. In contrast, the higher training loss and lower testing accuracy of NGCF reflect the practical difficulty to train such a heavy model it well. Note that in the figures we show the training process under the optimal hyper-parameter setting for both methods. Although increasing the learning rate of NGCF can decrease its training loss (even lower than that of LightGCN), the testing recall could not be improved, as lowering training loss in this way only finds trivial solution for NGCF. 
\end{itemize}


\subsection{Performance Comparison with State-of-the-Arts}\label{ss:exp-SOTA}
\begin{figure*}[t]
	\centering
	\subfigure{\includegraphics[width=0.24\textwidth]{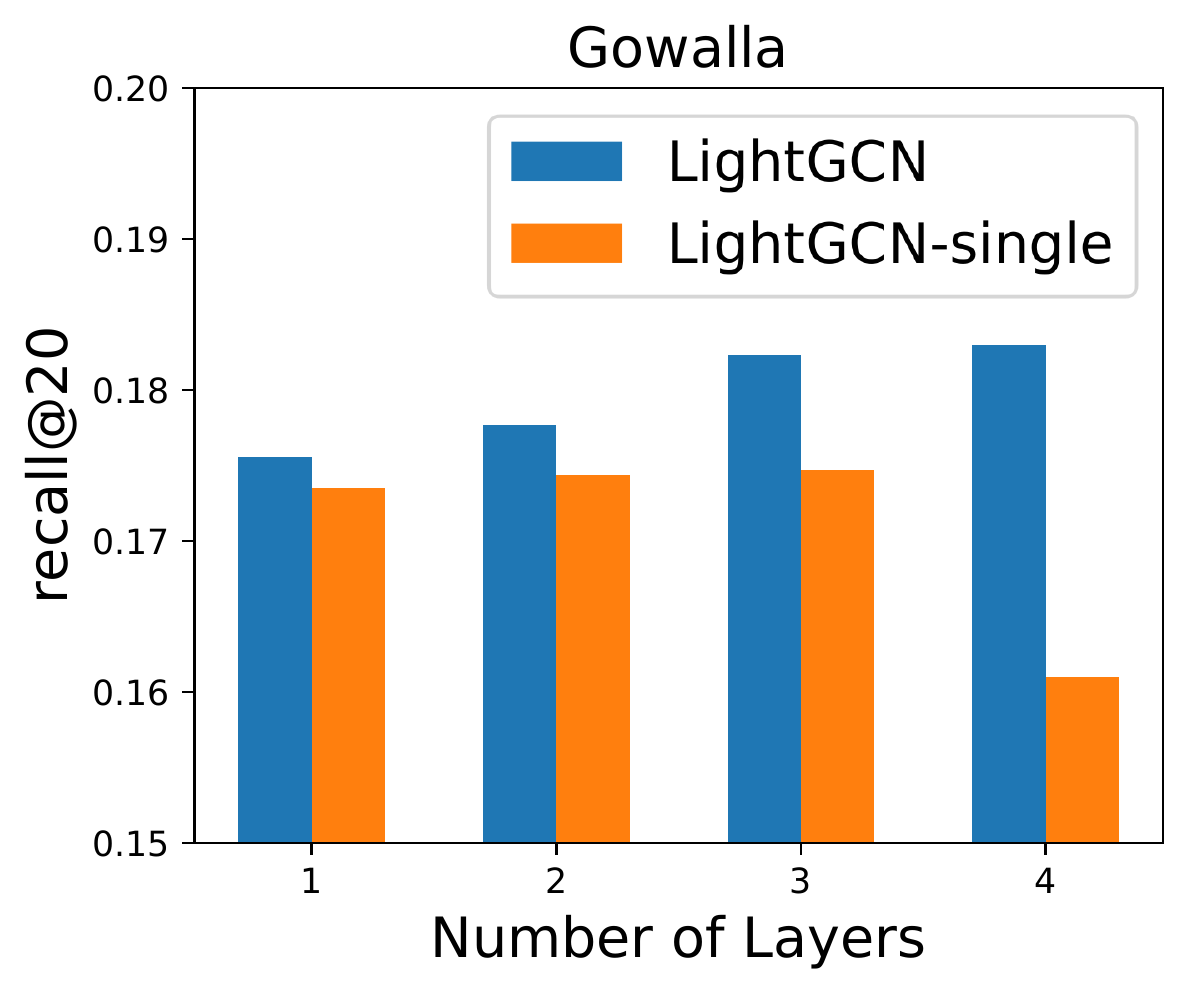}} 
	\subfigure{\includegraphics[width=0.24\textwidth]{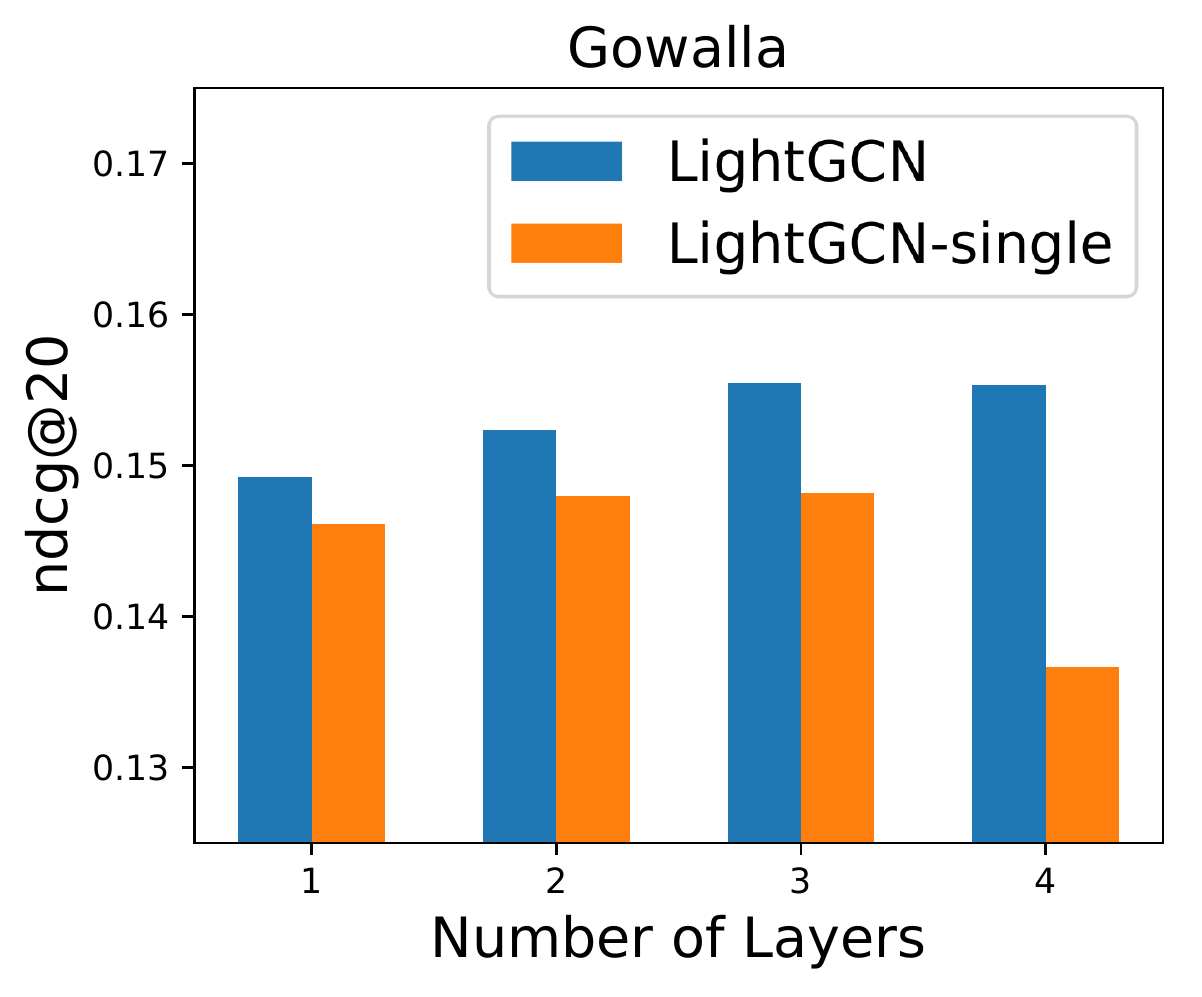}} 
	\subfigure{\includegraphics[width=0.24\textwidth]{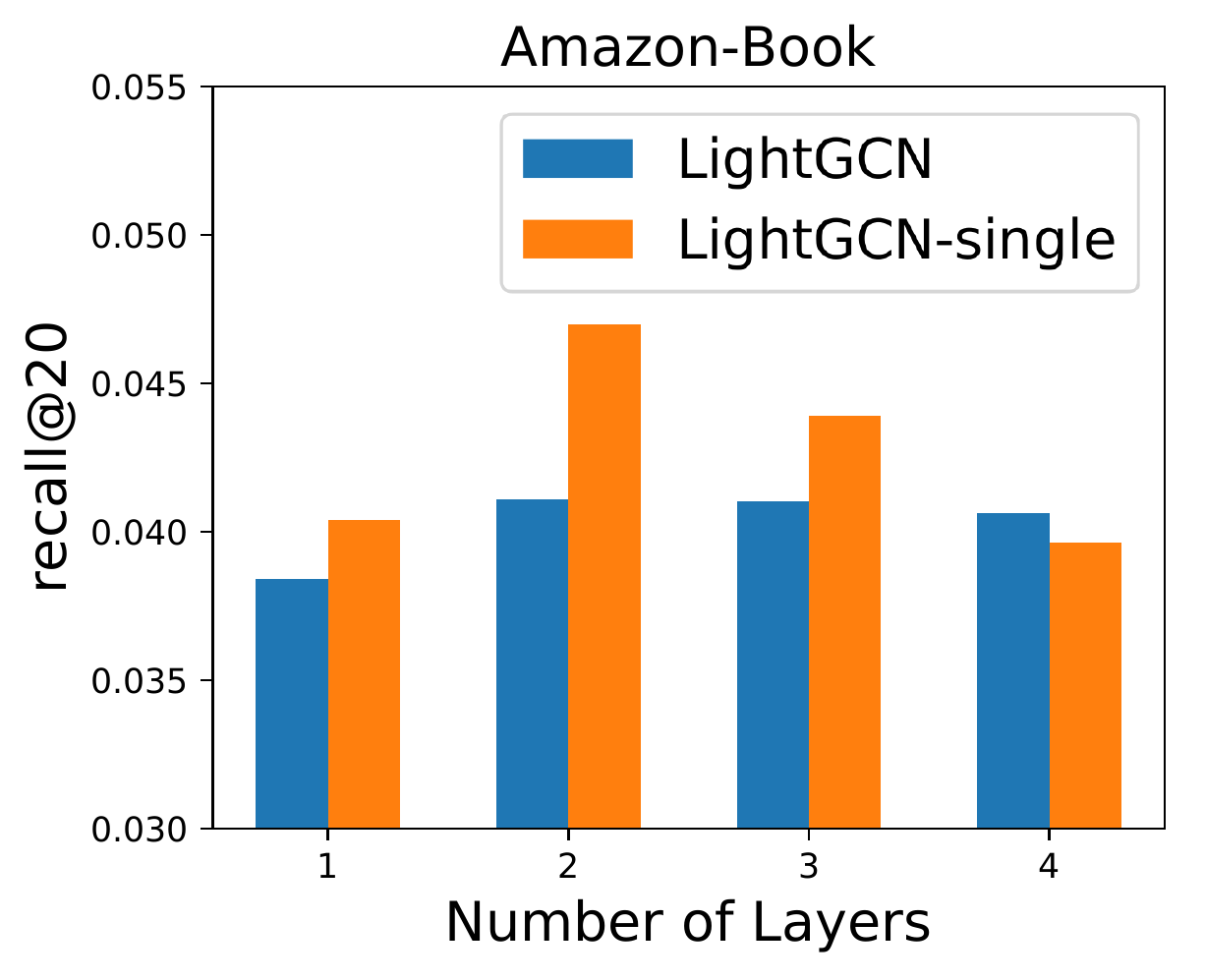}}
	\subfigure{\includegraphics[width=0.24\textwidth]{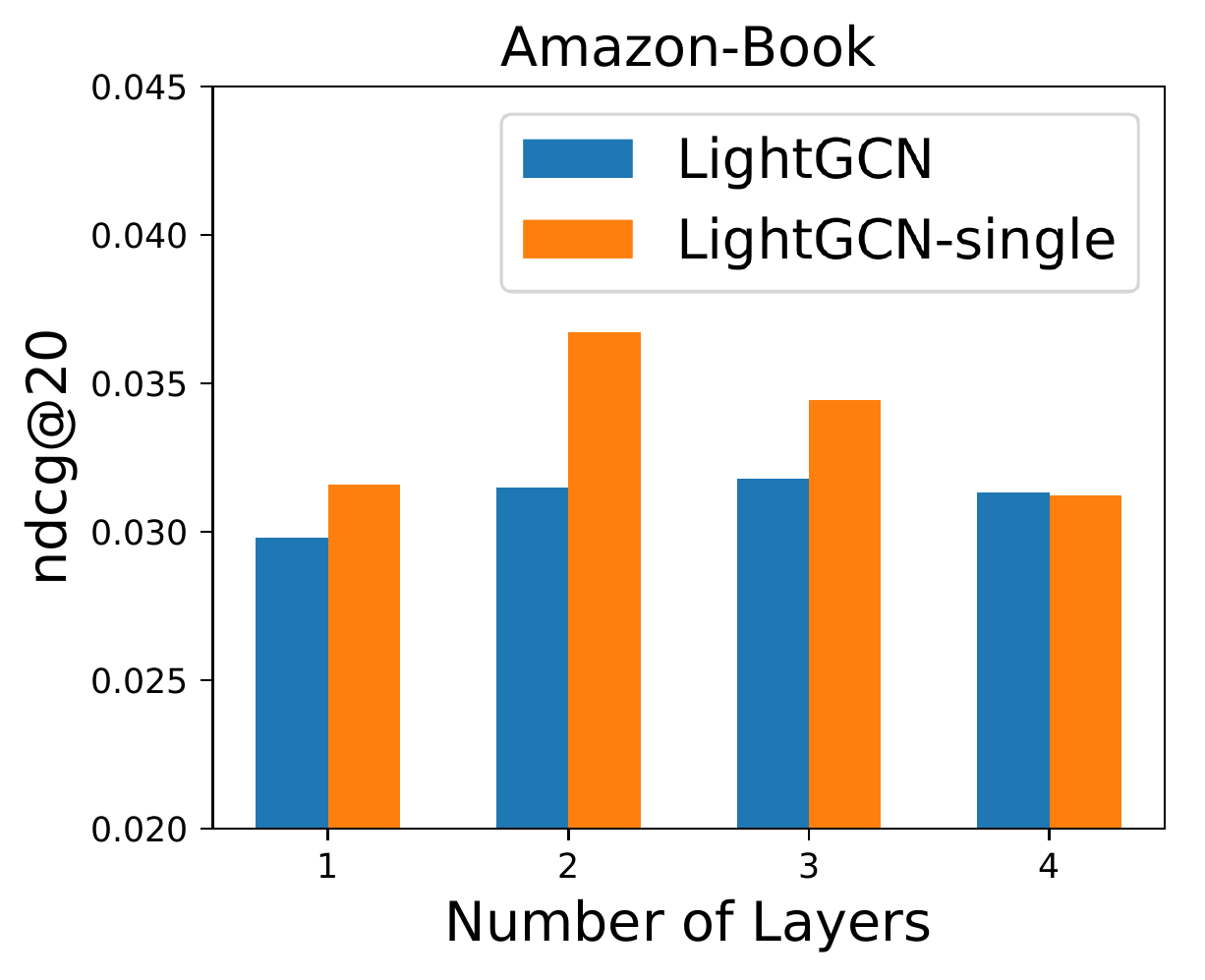}}
	\vspace{-15pt}
	\caption{Results of LightGCN and the variant that does not use layer combination (i.e., LightGCN-single) at different layers on Gowalla and Amazon-Book (results on Yelp2018 shows the same trend with Amazon-Book which are omitted for space).} \vspace{-5pt}
	\label{fig:layer-combination}
\end{figure*}

Table \ref{tab:overall} shows the performance comparison with competing methods. We show the best score we can obtain for each method. We can see that LightGCN consistently outperforms other methods on all three datasets, demonstrating its high effectiveness with simple yet reasonable designs. Note that LightGCN can be further improved by tuning the $\alpha_k$ (see Figure \ref{fig:layer-combination} for an evidence), while here we only use a uniform setting of $\frac{1}{K+1}$ to avoid over-tuning it. Among the baselines, Mult-VAE exhibits the strongest performance, which is better than GRMF and NGCF. 
The performance of GRMF is on a par with NGCF, being better than MF, which admits the utility of enforcing embedding smoothness with Laplacian regularizer. 
By adding normalization into the Laplacian regularizer, GRMF-norm betters than GRMF on Gowalla, while brings no benefits on Yelp2018 and Amazon-Book.

\begin{table}[h]
\caption{The comparison of overall performance among LightGCN and competing methods.}\vspace{-8pt}
\label{tab:overall}
\resizebox{0.48\textwidth}{!}{
\begin{tabular}{l|c c|c c|c c}
\hline
\textbf{Dataset} & \multicolumn{2}{c|}{\textbf{Gowalla}} & \multicolumn{2}{c|}{\textbf{Yelp2018}} & \multicolumn{2}{c}{\textbf{Amazon-Book}} \\ \hline
\textbf{Method} & \textbf{recall} & \textbf{ndcg} & \textbf{recall} & \textbf{ndcg} & \textbf{recall} & \textbf{ndcg} \\ \hline\hline
NGCF & 0.1570 & 0.1327 & 0.0579 & 0.0477 & 0.0344 & 0.0263 \\ \hline
Mult-VAE   & 0.1641	& 0.1335 & 0.0584 & 0.0450 &  0.0407 & 0.0315\\ \hline
GRMF & 0.1477 & 0.1205 & 0.0571 & 0.0462 & 0.0354 &
0.0270\\ 
GRMF-norm & 0.1557 & 0.1261 & 0.0561 & 0.0454
 & 0.0352 & 0.0269\\\hline
LightGCN & \textbf{0.1830} & \textbf{0.1554} & \textbf{0.0649} & \textbf{0.0530} & \textbf{0.0411} & \textbf{0.0315} \\\hline
\end{tabular}
}
\end{table}

\subsection{Ablation and Effectiveness Analyses}\label{ss:exp-ablation}
We perform ablation studies on LightGCN by showing how layer combination and symmetric sqrt normalization affect its performance. To justify the rationality of LightGCN as analyzed in Section \ref{ss:second-order}, we further investigate the effect of embedding smoothness --- the key reason of LightGCN's effectiveness. 


\subsubsection{Impact of Layer Combination} Figure \ref{fig:layer-combination} shows the results of LightGCN and its variant LightGCN-single that does not use layer combination (i.e., $\textbf{E}^{(K)}$ is used for final prediction for a $K$-layer LightGCN). We omit the results on Yelp2018 due to space limitation, which show similar trend with Amazon-Book. We have three main observations:
\begin{itemize}[leftmargin=*]
\item Focusing on LightGCN-single, we find that its performance first improves and then drops when the layer number increases from 1 to 4. The peak point is on layer 2 in most cases, while after that it drops quickly to the worst point of layer 4. This indicates that smoothing a node's embedding with its first-order and second-order neighbors is very useful for CF, but will suffer from over-smoothing issues when higher-order neighbors are used. 
\item Focusing on LightGCN, we find that its performance gradually improves with the increasing of layers. Even using 4 layers, LightGCN's performance is not degraded. This justifies the effectiveness of layer combination for addressing over-smoothing, as we have technically analyzed in Section~\ref{ss:APPNP} (relation with APPNP). 
\item Comparing the two methods, we find that LightGCN consistently outperforms LightGCN-single on Gowalla, but not on Amazon-Book and Yelp2018 (where the 2-layer LightGCN-single performs the best). Regarding this phenomenon, two points need to be noted before we draw conclusion: 1) LightGCN-single is special case of LightGCN that sets $\alpha_K$ to 1 and other $\alpha_k$ to 0; 2) we do not tune the $\alpha_k$ and simply set it as $\frac{1}{K+1}$ uniformly for LightGCN. As such, we can see the potential of further enhancing the performance of LightGCN by tuning $\alpha_k$. 
\end{itemize}

\begin{table}
\caption{Performance of the 3-layer LightGCN with different choices of normalization schemes in graph convolution.}\vspace{-8pt}
\label{tab:norm-3}
\resizebox{0.48\textwidth}{!}{
\begin{tabular}{l|c c|c c|c c}
\hline
\textbf{Dataset} & \multicolumn{2}{c|}{\textbf{Gowalla}} & \multicolumn{2}{c|}{\textbf{Yelp2018}} & \multicolumn{2}{c}{\textbf{Amazon-Book}} \\ \hline
\textbf{Method} & \textbf{recall} & \textbf{ndcg} & \textbf{recall} & \textbf{ndcg} & \textbf{recall} & \textbf{ndcg} \\ \hline\hline
LightGCN-$L_1$-L &  0.1724 & 0.1414
 & 0.0630 & 0.0511 & \textbf{0.0419} & \textbf{0.0320}\\ 
LightGCN-$L_1$-R & 0.1578 & 0.1348
 & 0.0587 & 0.0477 & 0.0334 & 0.0259\\ 
LightGCN-$L_1$ &0.159 & 0.1319
   &0.0573 & 0.0465 & 0.0361 & 0.0275\\ \hline
LightGCN-L    & 0.1589 & 0.1317 & 0.0619 & 0.0509
 & 0.0383 &	0.0299 \\ 
LightGCN-R   & 0.1420 & 0.1156 & 0.0521 & 0.0401
 & 0.0252 &	0.0196
 \\ \hline\hline
LightGCN & \textbf{0.1830} & \textbf{0.1554} & \textbf{0.0649} & \textbf{0.0530} & 0.0411 & 0.0315 \\ \hline 
\end{tabular}
}
\small{Method notation: -L means only the left-side norm is used, -R means only the right-side norm is used, and -$L_1$ means the $L_1$ norm is used.}
\end{table}
\subsubsection{Impact of Symmetric Sqrt Normalization}\label{ss:ablation-symmetric}
In LightGCN, we employ symmetric sqrt normalization $\frac{1}{\sqrt{|\mathcal{N}_u|}\sqrt{|\mathcal{N}_i|}}$ on each neighbor embedding when performing neighborhood aggregation (cf. Equation~(\ref{eq:LGC})). To study its rationality, we explore different choices here. We test the use of normalization only at the left side (i.e., the target node's coefficient) and the right side (i.e., the neighbor node's coefficient). 
We also test $L_1$ normalization, i.e.,  removing the square root. 
Note that if removing normalization, the training becomes numerically unstable and suffers from not-a-value (NAN) issues, so we do not show this setting. 
Table \ref{tab:norm-3} shows the results of the 3-layer LightGCN. 
We have the following observations:
\begin{itemize}[leftmargin=*]
    \item The best setting in general is using sqrt normalization at both sides (i.e., the current design of LightGCN). Removing either side will drop the performance largely.
    \item The second best setting is using $L_1$ normalization at the left side only (i.e., LightGCN-$L_1$-L). This is equivalent to normalize the adjacency matrix as a stochastic matrix by the in-degree. 
    \item Normalizing symmetrically on two sides is helpful for the sqrt normalization, but will degrade the performance of $L_1$ normalization. 
\end{itemize}

\subsubsection{Analysis of Embedding Smoothness} 
As we have analyzed in Section~\ref{ss:second-order}, a 2-layer LightGCN smooths a user's embedding based on the users that have overlap on her interacted items, and the smoothing strength between two users $c_{v\to u}$ is measured in Equation~(\ref{eq:c_vu}). We speculate that such smoothing of embeddings is the key reason of LightGCN's effectiveness. To verify this, we first define the smoothness of user embeddings as:
\begin{equation}
    S_U = \sum_{u=1}^M \sum_{v=1}^M c_{v\to u} (\frac{\textbf{e}_{u}}{||\textbf{e}_{u}||^2} - \frac{\textbf{e}_{v}}{||\textbf{e}_{v}||^2})^2, 
\end{equation}
where the $L_2$ norm on embeddings is used to eliminate the impact of the embedding's scale. Similarly we can obtained the definition for item embeddings. 
Table~\ref{tab:Smoothness} shows the smoothness of two models, matrix factorization (i.e., using the $\textbf{E}^{(0)}$ for model prediction) and the 2-layer LightGCN-single (i.e., using the $\textbf{E}^{(2)}$ for prediction). 
Note that the 2-layer LightGCN-single outperforms MF in recommendation accuracy by a large margin. As can be seen, the smoothness loss of LightGCN-single is much lower than that of MF. This indicates that by conducting light graph convolution, the embeddings become smoother and more suitable for recommendation.

\begin{table}[t]
\caption{Smoothness loss of the embeddings learned by LightGCN and MF (the lower the smoother).}\vspace{-8pt}
\label{tab:Smoothness}
\begin{tabular}{l|c|c|c}
\hline
\textbf{Dataset} & \multicolumn{1}{c|}{\textbf{Gowalla}} & \multicolumn{1}{c|}{\textbf{Yelp2018}} & \multicolumn{1}{c}{\textbf{Amazon-book}}\\ \hline
&\multicolumn{3}{c}{\textbf{Smoothness of User Embeddings}}\\ \hline
MF & 15449.3 & 16258.2 & 38034.2  \\ \hline
LightGCN-single & 12872.7 & 10091.7 & 32191.1 \\ \hline
&\multicolumn{3}{c}{\textbf{Smoothness of Item Embeddings}}\\ \hline
MF & 12106.7 & 16632.1 & 28307.9  \\ \hline
LightGCN-single & 5829.0 & 6459.8 & 16866.0 \\ \hline
\end{tabular}
\vspace{-5pt}
\end{table}


\subsection{Hyper-parameter Studies}\label{ss:exp-hyper}
When applying LightGCN to a new dataset, besides the standard hyper-parameter learning rate, the most important hyper-parameter to tune is the $L_2$ regularization coefficient $\lambda$. 
Here we investigate the performance change of LightGCN w.r.t. $\lambda$. 

\begin{figure}[t]
	\centering
	\subfigure{\includegraphics[width=0.23\textwidth]{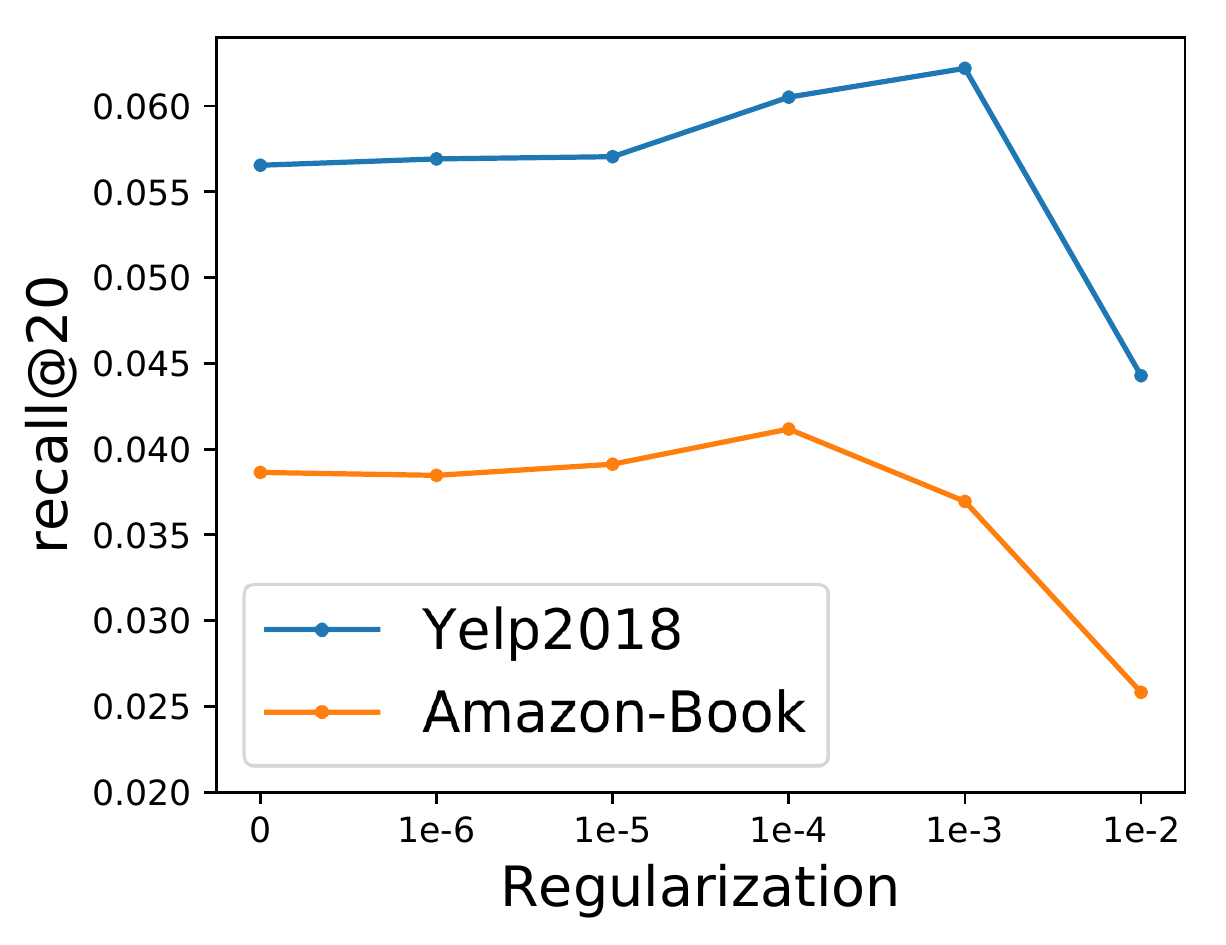}} 
	\subfigure{\includegraphics[width=0.23\textwidth]{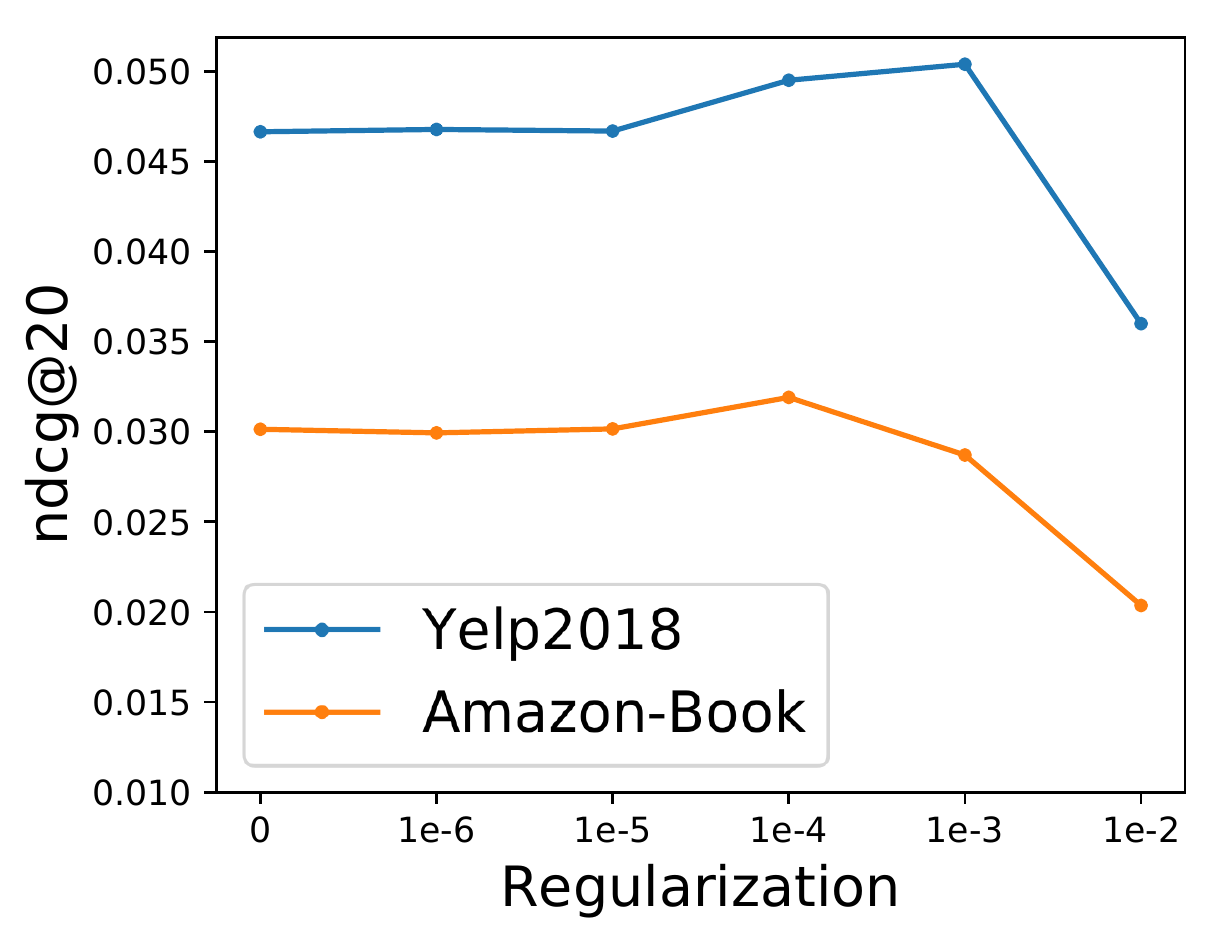}} 
	\vspace{-15pt}
	\caption{Performance of 2-layer LightGCN w.r.t. different regularization coefficient $\lambda$ on Yelp and Amazon-Book.} \vspace{-10pt}
	\label{fig:reg}
\end{figure}

As shown in Figure \ref{fig:reg},
LightGCN is relatively insensitive to $\lambda$ --- even when $\lambda$ sets to 0, LightGCN is better than NGCF, which additionally uses dropout to prevent overfitting\footnote{Note that Gowalla shows the same trend with Amazon-Book, so its curves are not shown to better highlight the trend of Yelp2018 and Amazon-Book.}. 
This shows that LightGCN is less prone to overfitting --- since the only trainable parameters in LightGCN are ID embeddings of the 0-th layer, the whole model is easy to train and to regularize. The optimal value for Yelp2018, Amazon-Book, and Gowalla is $1e^{-3}$, $1e^{-4}$, and  $1e^{-4}$, respectively. When $\lambda$ is larger than $1e^{-3}$, the performance drops quickly, which indicates that too strong regularization will negatively affect model normal training and is not encouraged. 

\section{Related Work}\label{sec:related}

\subsection{Collaborative Filtering}
Collaborative Filtering (CF) is a prevalent technique in modern recommender systems~\cite{YoutubeRS,PinSage}.
One common paradigm of CF model is to parameterize users and items as embeddings, and learn the embedding parameters by reconstructing historical user-item interactions.
For example, earlier CF models like matrix factorization (MF)~\cite{MF,BPRMF} project the ID of a user (or an item) into an embedding vector. The recent neural recommender models like NCF~\cite{NCF} and LRML~\cite{tay2018latent} use the same embedding component, while enhance the interaction modeling with neural networks. 



Beyond merely using ID information, another type of CF methods considers historical items as the pre-existing features of a user, towards better user representations.
For example,  FISM~\cite{FISM} and SVD++~\cite{SVD++} use the weighted average of the ID embeddings of historical items as the target user's embedding. 
Recently, researchers realize that historical items have different contributions to shape personal interest.
Towards this end, attention mechanisms are introduced to capture the varying contributions, such as ACF~\cite{ACF} and NAIS~\cite{NAIS}, to automatically learn the importance of each historical item. 
When revisiting historical interactions as a user-item bipartite graph, the performance improvements can be attributed to the encoding of local neighborhood --- one-hop neighbors --- that improves the embedding learning.

\subsection{Graph Methods for Recommendation}
Another relevant research line is exploiting the user-item graph structure for recommendation.
Prior efforts like ItemRank~\cite{ItemRank}, 
use the label propagation mechanism to directly propagate user preference scores over the graph, \ie encouraging connected nodes to have similar labels.
Recently emerged graph neural networks (GNNs) shine a light on modeling graph structure, especially high-hop neighbors, to guide the embedding learning~\cite{GCN,GraphSAGE}.
Early studies define graph convolution on the spectral domain, such as Laplacian eigen-decomposition~\cite{DBLP:journals/corr/BrunaZSL13} and Chebyshev polynomials~\cite{FirstGCN}, which are computationally expensive.
Later on, GraphSage~\cite{GraphSAGE} and GCN~\cite{GCN} re-define graph convolution in the spatial domain, i.e., aggregating the embeddings of neighbors to refine the target node's embedding. Owing to its interpretability and efficiency, it quickly becomes a prevalent formulation of GNNs and is being widely used~\cite{DeepInf,Feng2019TOIS,zhao2019cross}.
Motivated by the strength of graph convolution, recent efforts like NGCF~\cite{NGCF}, GC-MC~\cite{GC-MC}, and PinSage~\cite{PinSage} adapt GCN to the user-item interaction graph, capturing CF signals in high-hop neighbors for recommendation.

It is worth mentioning that several recent efforts provide deep insights into GNNs~\cite{DeepInsights,ICLR19-APPNP,SGCN}, which inspire us developing LightGCN. Particularly, Wu et al.~\cite{SGCN} argues the unnecessary complexity of GCN, developing a simplified GCN (SGCN) model by removing nonlinearities and collapsing multiple weight matrices into one. 
One main difference is that LightGCN and SGCN are developed for different tasks, thus the rationality of model simplification is different. Specifically, SGCN is for node classification, performing simplification for model interpretability and efficiency. In contrast, LightGCN is on collaborative filtering (CF), where each node has an ID feature only. Thus, we do simplification for a stronger reason: nonlinearity and weight matrices are useless for CF, and even hurt model training. For node classification accuracy, SGCN is on par with (sometimes weaker than) GCN. While for CF accuracy, LightGCN outperforms GCN by a large margin (over 15\% improvement over NGCF). 
Lastly, another work conducted in the same time~\cite{LR-GCCF} also finds that the nonlinearity is unnecessary in NGCF and develops linear GCN model for CF. In contrast, our LightGCN makes one step further --- we remove all redundant parameters and retain only the ID embeddings, making the model as simple as MF.

\section{Conclusion and Future Work} \label{sec:conclusion}
In this work, we argued the unnecessarily complicated design of GCNs for collaborative filtering, and performed empirical studies to justify this argument. 
We proposed LightGCN which consists of two essential components --- light graph convolution and layer combination. 
In light graph convolution, we 
discard feature transformation and nonlinear activation --- two standard operations in GCNs but inevitably increase the training difficulty. In layer combination, we construct a node's final embedding as the weighted sum of its embeddings on all layers, which is proved to subsume the effect of self-connections and is helpful to control oversmoothing. 
We conduct experiments to demonstrate the strengths of LightGCN in being simple: easier to be trained, better generalization ability, and more effective. 

We believe the insights of LightGCN are inspirational to future developments of recommender models. With the prevalence of linked graph data in real applications, graph-based models are becoming increasingly important in recommendation;
by explicitly exploiting the relations among entities in the predictive model, they are advantageous to traditional supervised learning scheme like factorization machines~\cite{FM,NFM} that model the relations implicitly. 
For example, a recent trend is to exploit auxiliary information such as item knowledge graph~\cite{KGAT}, social network~\cite{GCNSocial} and multimedia content~\cite{MMGCN} for recommendation, where GCNs have set up the new state-of-the-art. 
However, these models may also suffer from the similar issues of NGCF since the user-item interaction graph is also modeled by same neural operations that may be unnecessary. We plan to explore the idea of LightGCN in these models. Another future direction is to personalize the layer combination weights $\alpha_k$, so as to enable adaptive-order smoothing for different users (e.g., sparse users may require more signal from higher-order neighbors while active users require less). Lastly, we will explore further the strengths of LightGCN's simplicity, studying whether fast solution exists for non-sampling regression loss~\cite{he2019fast} and streaming it for online industrial scenarios. 

\noindent\textbf{Acknowledgement}. The authors thank Bin Wu, Jianbai Ye, and Yingxin Wu for contributing to the implementation and improvement of LightGCN. This work is supported by the National Natural Science Foundation of China (61972372, U19A2079, 61725203).

\bibliographystyle{ACM-Reference-Format}
\balance

\bibliography{ms}


\begin{thebibliography}{48}


\ifx \showCODEN    \undefined \def \showCODEN     #1{\unskip}     \fi
\ifx \showDOI      \undefined \def \showDOI       #1{#1}\fi
\ifx \showISBNx    \undefined \def \showISBNx     #1{\unskip}     \fi
\ifx \showISBNxiii \undefined \def \showISBNxiii  #1{\unskip}     \fi
\ifx \showISSN     \undefined \def \showISSN      #1{\unskip}     \fi
\ifx \showLCCN     \undefined \def \showLCCN      #1{\unskip}     \fi
\ifx \shownote     \undefined \def \shownote      #1{#1}          \fi
\ifx \showarticletitle \undefined \def \showarticletitle #1{#1}   \fi
\ifx \showURL      \undefined \def \showURL       {\relax}        \fi
\providecommand\bibfield[2]{#2}
\providecommand\bibinfo[2]{#2}
\providecommand\natexlab[1]{#1}
\providecommand\showeprint[2][]{arXiv:#2}

\bibitem[\protect\citeauthoryear{Bruna, Zaremba, Szlam, and LeCun}{Bruna
  et~al\mbox{.}}{2014}]%
        {DBLP:journals/corr/BrunaZSL13}
\bibfield{author}{\bibinfo{person}{Joan Bruna}, \bibinfo{person}{Wojciech
  Zaremba}, \bibinfo{person}{Arthur Szlam}, {and} \bibinfo{person}{Yann
  LeCun}.} \bibinfo{year}{2014}\natexlab{}.
\newblock \showarticletitle{Spectral Networks and Locally Connected Networks on
  Graphs}. In \bibinfo{booktitle}{\emph{ICLR}}.
\newblock


\bibitem[\protect\citeauthoryear{Chen, Wang, Tsai, and Yang}{Chen
  et~al\mbox{.}}{2019b}]%
        {CSE}
\bibfield{author}{\bibinfo{person}{Chih{-}Ming Chen},
  \bibinfo{person}{Chuan{-}Ju Wang}, \bibinfo{person}{Ming{-}Feng Tsai}, {and}
  \bibinfo{person}{Yi{-}Hsuan Yang}.} \bibinfo{year}{2019}\natexlab{b}.
\newblock \showarticletitle{Collaborative Similarity Embedding for Recommender
  Systems}. In \bibinfo{booktitle}{\emph{WWW}}. \bibinfo{pages}{2637--2643}.
\newblock


\bibitem[\protect\citeauthoryear{Chen, Zhang, He, Nie, Liu, and Chua}{Chen
  et~al\mbox{.}}{2017}]%
        {ACF}
\bibfield{author}{\bibinfo{person}{Jingyuan Chen}, \bibinfo{person}{Hanwang
  Zhang}, \bibinfo{person}{Xiangnan He}, \bibinfo{person}{Liqiang Nie},
  \bibinfo{person}{Wei Liu}, {and} \bibinfo{person}{Tat{-}Seng Chua}.}
  \bibinfo{year}{2017}\natexlab{}.
\newblock \showarticletitle{Attentive Collaborative Filtering: Multimedia
  Recommendation with Item- and Component-Level Attention}. In
  \bibinfo{booktitle}{\emph{{SIGIR}}}. \bibinfo{pages}{335--344}.
\newblock


\bibitem[\protect\citeauthoryear{Chen, Wu, Hong, Zhang, and Wang}{Chen
  et~al\mbox{.}}{2020}]%
        {LR-GCCF}
\bibfield{author}{\bibinfo{person}{Lei Chen}, \bibinfo{person}{Le Wu},
  \bibinfo{person}{Richang Hong}, \bibinfo{person}{Kun Zhang}, {and}
  \bibinfo{person}{Meng Wang}.} \bibinfo{year}{2020}\natexlab{}.
\newblock \showarticletitle{Revisiting Graph based Collaborative Filtering: A
  Linear Residual Graph Convolutional Network Approach}. In
  \bibinfo{booktitle}{\emph{AAAI}}.
\newblock


\bibitem[\protect\citeauthoryear{Chen, Chen, He, Gao, Li, Lou, and Wang}{Chen
  et~al\mbox{.}}{2019a}]%
        {lambdaOpt}
\bibfield{author}{\bibinfo{person}{Yihong Chen}, \bibinfo{person}{Bei Chen},
  \bibinfo{person}{Xiangnan He}, \bibinfo{person}{Chen Gao},
  \bibinfo{person}{Yong Li}, \bibinfo{person}{Jian-Guang Lou}, {and}
  \bibinfo{person}{Yue Wang}.} \bibinfo{year}{2019}\natexlab{a}.
\newblock \showarticletitle{{\(\lambda\)}Opt: Learn to Regularize Recommender
  Models in Finer Levels}. In \bibinfo{booktitle}{\emph{KDD}}.
  \bibinfo{pages}{978--986}.
\newblock


\bibitem[\protect\citeauthoryear{Cheng, Ding, Zhu, and Kankanhalli}{Cheng
  et~al\mbox{.}}{2018}]%
        {DBLP:conf/www/ChengDZK18}
\bibfield{author}{\bibinfo{person}{Zhiyong Cheng}, \bibinfo{person}{Ying Ding},
  \bibinfo{person}{Lei Zhu}, {and} \bibinfo{person}{Mohan~S. Kankanhalli}.}
  \bibinfo{year}{2018}\natexlab{}.
\newblock \showarticletitle{Aspect-Aware Latent Factor Model: Rating Prediction
  with Ratings and Reviews}. In \bibinfo{booktitle}{\emph{{WWW}}}.
  \bibinfo{pages}{639--648}.
\newblock


\bibitem[\protect\citeauthoryear{Covington, Adams, and Sargin}{Covington
  et~al\mbox{.}}{2016}]%
        {YoutubeRS}
\bibfield{author}{\bibinfo{person}{Paul Covington}, \bibinfo{person}{Jay
  Adams}, {and} \bibinfo{person}{Emre Sargin}.}
  \bibinfo{year}{2016}\natexlab{}.
\newblock \showarticletitle{Deep Neural Networks for YouTube Recommendations}.
  In \bibinfo{booktitle}{\emph{RecSys}}. \bibinfo{pages}{191--198}.
\newblock


\bibitem[\protect\citeauthoryear{Defferrard, Bresson, and
  Vandergheynst}{Defferrard et~al\mbox{.}}{2016}]%
        {FirstGCN}
\bibfield{author}{\bibinfo{person}{Micha{\"{e}}l Defferrard},
  \bibinfo{person}{Xavier Bresson}, {and} \bibinfo{person}{Pierre
  Vandergheynst}.} \bibinfo{year}{2016}\natexlab{}.
\newblock \showarticletitle{Convolutional Neural Networks on Graphs with Fast
  Localized Spectral Filtering}. In \bibinfo{booktitle}{\emph{{NeurIPS}}}.
  \bibinfo{pages}{3837--3845}.
\newblock


\bibitem[\protect\citeauthoryear{Ding, Quan, He, Li, and Jin}{Ding
  et~al\mbox{.}}{2019}]%
        {Ding2019IJCAI}
\bibfield{author}{\bibinfo{person}{Jingtao Ding}, \bibinfo{person}{Yuhan Quan},
  \bibinfo{person}{Xiangnan He}, \bibinfo{person}{Yong Li}, {and}
  \bibinfo{person}{Depeng Jin}.} \bibinfo{year}{2019}\natexlab{}.
\newblock \showarticletitle{Reinforced Negative Sampling for Recommendation
  with Exposure Data}. In \bibinfo{booktitle}{\emph{IJCAI}}.
  \bibinfo{pages}{2230--2236}.
\newblock


\bibitem[\protect\citeauthoryear{Ebesu, Shen, and Fang}{Ebesu
  et~al\mbox{.}}{2018}]%
        {CMN}
\bibfield{author}{\bibinfo{person}{Travis Ebesu}, \bibinfo{person}{Bin Shen},
  {and} \bibinfo{person}{Yi Fang}.} \bibinfo{year}{2018}\natexlab{}.
\newblock \showarticletitle{Collaborative Memory Network for Recommendation
  Systems}. In \bibinfo{booktitle}{\emph{{SIGIR}}}. \bibinfo{pages}{515--524}.
\newblock


\bibitem[\protect\citeauthoryear{Feng, He, Wang, Luo, Liu, and Chua}{Feng
  et~al\mbox{.}}{2019}]%
        {Feng2019TOIS}
\bibfield{author}{\bibinfo{person}{Fuli Feng}, \bibinfo{person}{Xiangnan He},
  \bibinfo{person}{Xiang Wang}, \bibinfo{person}{Cheng Luo},
  \bibinfo{person}{Yiqun Liu}, {and} \bibinfo{person}{Tat-Seng Chua}.}
  \bibinfo{year}{2019}\natexlab{}.
\newblock \showarticletitle{Temporal Relational Ranking for Stock Prediction}.
\newblock \bibinfo{journal}{\emph{TOIS}} \bibinfo{volume}{37},
  \bibinfo{number}{2} (\bibinfo{year}{2019}), \bibinfo{pages}{27:1--27:30}.
\newblock


\bibitem[\protect\citeauthoryear{Glorot and Bengio}{Glorot and Bengio}{2010}]%
        {Xarvier}
\bibfield{author}{\bibinfo{person}{Xavier Glorot} {and} \bibinfo{person}{Yoshua
  Bengio}.} \bibinfo{year}{2010}\natexlab{}.
\newblock \showarticletitle{Understanding the difficulty of training deep
  feedforward neural networks}. In \bibinfo{booktitle}{\emph{{AISTATS}}}.
  \bibinfo{pages}{249--256}.
\newblock


\bibitem[\protect\citeauthoryear{Gori and Pucci}{Gori and Pucci}{2007}]%
        {ItemRank}
\bibfield{author}{\bibinfo{person}{Marco Gori} {and} \bibinfo{person}{Augusto
  Pucci}.} \bibinfo{year}{2007}\natexlab{}.
\newblock \showarticletitle{ItemRank: {A} Random-Walk Based Scoring Algorithm
  for Recommender Engines}. In \bibinfo{booktitle}{\emph{{IJCAI}}}.
  \bibinfo{pages}{2766--2771}.
\newblock


\bibitem[\protect\citeauthoryear{Hamilton, Ying, and Leskovec}{Hamilton
  et~al\mbox{.}}{2017}]%
        {GraphSAGE}
\bibfield{author}{\bibinfo{person}{William~L. Hamilton},
  \bibinfo{person}{Zhitao Ying}, {and} \bibinfo{person}{Jure Leskovec}.}
  \bibinfo{year}{2017}\natexlab{}.
\newblock \showarticletitle{Inductive Representation Learning on Large Graphs}.
  In \bibinfo{booktitle}{\emph{{NeurIPS}}}. \bibinfo{pages}{1025--1035}.
\newblock


\bibitem[\protect\citeauthoryear{Haveliwala}{Haveliwala}{2002}]%
        {haveliwala2002topic}
\bibfield{author}{\bibinfo{person}{Taher~H Haveliwala}.}
  \bibinfo{year}{2002}\natexlab{}.
\newblock \showarticletitle{Topic-sensitive pagerank}. In
  \bibinfo{booktitle}{\emph{WWW}}. \bibinfo{pages}{517--526}.
\newblock


\bibitem[\protect\citeauthoryear{He, Zhang, Ren, and Sun}{He
  et~al\mbox{.}}{2016}]%
        {ResNet}
\bibfield{author}{\bibinfo{person}{Kaiming He}, \bibinfo{person}{Xiangyu
  Zhang}, \bibinfo{person}{Shaoqing Ren}, {and} \bibinfo{person}{Jian Sun}.}
  \bibinfo{year}{2016}\natexlab{}.
\newblock \showarticletitle{Deep residual learning for image recognition}. In
  \bibinfo{booktitle}{\emph{CVPR}}. \bibinfo{pages}{770--778}.
\newblock


\bibitem[\protect\citeauthoryear{He and Chua}{He and Chua}{2017}]%
        {NFM}
\bibfield{author}{\bibinfo{person}{Xiangnan He} {and}
  \bibinfo{person}{Tat{-}Seng Chua}.} \bibinfo{year}{2017}\natexlab{}.
\newblock \showarticletitle{Neural Factorization Machines for Sparse Predictive
  Analytics}. In \bibinfo{booktitle}{\emph{{SIGIR}}}.
  \bibinfo{pages}{355--364}.
\newblock


\bibitem[\protect\citeauthoryear{He, He, Song, Liu, Jiang, and Chua}{He
  et~al\mbox{.}}{2018}]%
        {NAIS}
\bibfield{author}{\bibinfo{person}{Xiangnan He}, \bibinfo{person}{Zhankui He},
  \bibinfo{person}{Jingkuan Song}, \bibinfo{person}{Zhenguang Liu},
  \bibinfo{person}{Yu{-}Gang Jiang}, {and} \bibinfo{person}{Tat{-}Seng Chua}.}
  \bibinfo{year}{2018}\natexlab{}.
\newblock \showarticletitle{{NAIS:} Neural Attentive Item Similarity Model for
  Recommendation}.
\newblock \bibinfo{journal}{\emph{TKDE}} \bibinfo{volume}{30},
  \bibinfo{number}{12} (\bibinfo{year}{2018}), \bibinfo{pages}{2354--2366}.
\newblock


\bibitem[\protect\citeauthoryear{He, Liao, Zhang, Nie, Hu, and Chua}{He
  et~al\mbox{.}}{2017}]%
        {NCF}
\bibfield{author}{\bibinfo{person}{Xiangnan He}, \bibinfo{person}{Lizi Liao},
  \bibinfo{person}{Hanwang Zhang}, \bibinfo{person}{Liqiang Nie},
  \bibinfo{person}{Xia Hu}, {and} \bibinfo{person}{Tat{-}Seng Chua}.}
  \bibinfo{year}{2017}\natexlab{}.
\newblock \showarticletitle{Neural Collaborative Filtering}. In
  \bibinfo{booktitle}{\emph{{WWW}}}. \bibinfo{pages}{173--182}.
\newblock


\bibitem[\protect\citeauthoryear{He, Tang, Du, Hong, Ren, and Chua}{He
  et~al\mbox{.}}{2019}]%
        {he2019fast}
\bibfield{author}{\bibinfo{person}{Xiangnan He}, \bibinfo{person}{Jinhui Tang},
  \bibinfo{person}{Xiaoyu Du}, \bibinfo{person}{Richang Hong},
  \bibinfo{person}{Tongwei Ren}, {and} \bibinfo{person}{Tat-Seng Chua}.}
  \bibinfo{year}{2019}\natexlab{}.
\newblock \showarticletitle{Fast Matrix Factorization with Nonuniform Weights
  on Missing Data}.
\newblock \bibinfo{journal}{\emph{TNNLS}} (\bibinfo{year}{2019}).
\newblock


\bibitem[\protect\citeauthoryear{Kabbur, Ning, and Karypis}{Kabbur
  et~al\mbox{.}}{2013}]%
        {FISM}
\bibfield{author}{\bibinfo{person}{Santosh Kabbur}, \bibinfo{person}{Xia Ning},
  {and} \bibinfo{person}{George Karypis}.} \bibinfo{year}{2013}\natexlab{}.
\newblock \showarticletitle{{FISM:} factored item similarity models for top-N
  recommender systems}. In \bibinfo{booktitle}{\emph{{KDD}}}.
  \bibinfo{pages}{659--667}.
\newblock


\bibitem[\protect\citeauthoryear{Kingma and Ba}{Kingma and Ba}{2015}]%
        {Adam}
\bibfield{author}{\bibinfo{person}{Diederik~P. Kingma} {and}
  \bibinfo{person}{Jimmy Ba}.} \bibinfo{year}{2015}\natexlab{}.
\newblock \showarticletitle{Adam: {A} Method for Stochastic Optimization}. In
  \bibinfo{booktitle}{\emph{{ICLR}}}.
\newblock


\bibitem[\protect\citeauthoryear{Kipf and Welling}{Kipf and Welling}{2017}]%
        {GCN}
\bibfield{author}{\bibinfo{person}{Thomas~N. Kipf} {and} \bibinfo{person}{Max
  Welling}.} \bibinfo{year}{2017}\natexlab{}.
\newblock \showarticletitle{Semi-Supervised Classification with Graph
  Convolutional Networks}. In \bibinfo{booktitle}{\emph{{ICLR}}}.
\newblock


\bibitem[\protect\citeauthoryear{Klicpera, Bojchevski, and
  G{\"u}nnemann}{Klicpera et~al\mbox{.}}{2019}]%
        {ICLR19-APPNP}
\bibfield{author}{\bibinfo{person}{Johannes Klicpera},
  \bibinfo{person}{Aleksandar Bojchevski}, {and} \bibinfo{person}{Stephan
  G{\"u}nnemann}.} \bibinfo{year}{2019}\natexlab{}.
\newblock \showarticletitle{Predict then propagate: Graph neural networks meet
  personalized pagerank}. In \bibinfo{booktitle}{\emph{{ICLR}}}.
\newblock


\bibitem[\protect\citeauthoryear{Koren}{Koren}{2008}]%
        {SVD++}
\bibfield{author}{\bibinfo{person}{Yehuda Koren}.}
  \bibinfo{year}{2008}\natexlab{}.
\newblock \showarticletitle{Factorization meets the neighborhood: a
  multifaceted collaborative filtering model}. In
  \bibinfo{booktitle}{\emph{{KDD}}}. \bibinfo{pages}{426--434}.
\newblock


\bibitem[\protect\citeauthoryear{Koren, Bell, and Volinsky}{Koren
  et~al\mbox{.}}{2009}]%
        {MF}
\bibfield{author}{\bibinfo{person}{Yehuda Koren}, \bibinfo{person}{Robert~M.
  Bell}, {and} \bibinfo{person}{Chris Volinsky}.}
  \bibinfo{year}{2009}\natexlab{}.
\newblock \showarticletitle{Matrix Factorization Techniques for Recommender
  Systems}.
\newblock \bibinfo{journal}{\emph{{IEEE} Computer}} \bibinfo{volume}{42},
  \bibinfo{number}{8} (\bibinfo{year}{2009}), \bibinfo{pages}{30--37}.
\newblock


\bibitem[\protect\citeauthoryear{Li, Han, and Wu}{Li et~al\mbox{.}}{2018}]%
        {DeepInsights}
\bibfield{author}{\bibinfo{person}{Qimai Li}, \bibinfo{person}{Zhichao Han},
  {and} \bibinfo{person}{Xiao{-}Ming Wu}.} \bibinfo{year}{2018}\natexlab{}.
\newblock \showarticletitle{Deeper Insights Into Graph Convolutional Networks
  for Semi-Supervised Learning}. In \bibinfo{booktitle}{\emph{AAAI}}.
  \bibinfo{pages}{3538--3545}.
\newblock


\bibitem[\protect\citeauthoryear{Liang, Krishnan, Hoffman, and Jebara}{Liang
  et~al\mbox{.}}{2018}]%
        {VACF}
\bibfield{author}{\bibinfo{person}{Dawen Liang}, \bibinfo{person}{Rahul~G.
  Krishnan}, \bibinfo{person}{Matthew~D. Hoffman}, {and} \bibinfo{person}{Tony
  Jebara}.} \bibinfo{year}{2018}\natexlab{}.
\newblock \showarticletitle{Variational Autoencoders for Collaborative
  Filtering}. In \bibinfo{booktitle}{\emph{{WWW}}}. \bibinfo{pages}{689--698}.
\newblock


\bibitem[\protect\citeauthoryear{Qiu, Tang, Ma, Dong, Wang, and Tang}{Qiu
  et~al\mbox{.}}{2018}]%
        {DeepInf}
\bibfield{author}{\bibinfo{person}{Jiezhong Qiu}, \bibinfo{person}{Jian Tang},
  \bibinfo{person}{Hao Ma}, \bibinfo{person}{Yuxiao Dong},
  \bibinfo{person}{Kuansan Wang}, {and} \bibinfo{person}{Jie Tang}.}
  \bibinfo{year}{2018}\natexlab{}.
\newblock \showarticletitle{DeepInf: Social Influence Prediction with Deep
  Learning}. In \bibinfo{booktitle}{\emph{{KDD}}}. \bibinfo{pages}{2110--2119}.
\newblock


\bibitem[\protect\citeauthoryear{Rao, Yu, Ravikumar, and Dhillon}{Rao
  et~al\mbox{.}}{2015}]%
        {rao2015collaborative}
\bibfield{author}{\bibinfo{person}{Nikhil Rao}, \bibinfo{person}{Hsiang-Fu Yu},
  \bibinfo{person}{Pradeep~K Ravikumar}, {and} \bibinfo{person}{Inderjit~S
  Dhillon}.} \bibinfo{year}{2015}\natexlab{}.
\newblock \showarticletitle{Collaborative filtering with graph information:
  Consistency and scalable methods}. In \bibinfo{booktitle}{\emph{{NIPS}}}.
  \bibinfo{pages}{2107--2115}.
\newblock


\bibitem[\protect\citeauthoryear{Rendle and Freudenthaler}{Rendle and
  Freudenthaler}{2014}]%
        {rendle2014improving}
\bibfield{author}{\bibinfo{person}{Steffen Rendle} {and}
  \bibinfo{person}{Christoph Freudenthaler}.} \bibinfo{year}{2014}\natexlab{}.
\newblock \showarticletitle{Improving pairwise learning for item recommendation
  from implicit feedback}. In \bibinfo{booktitle}{\emph{WSDM}}.
  \bibinfo{pages}{273--282}.
\newblock


\bibitem[\protect\citeauthoryear{Rendle, Freudenthaler, Gantner, and
  Schmidt{-}Thieme}{Rendle et~al\mbox{.}}{2009}]%
        {BPRMF}
\bibfield{author}{\bibinfo{person}{Steffen Rendle}, \bibinfo{person}{Christoph
  Freudenthaler}, \bibinfo{person}{Zeno Gantner}, {and} \bibinfo{person}{Lars
  Schmidt{-}Thieme}.} \bibinfo{year}{2009}\natexlab{}.
\newblock \showarticletitle{{BPR:} Bayesian Personalized Ranking from Implicit
  Feedback}. In \bibinfo{booktitle}{\emph{{UAI}}}. \bibinfo{pages}{452--461}.
\newblock


\bibitem[\protect\citeauthoryear{Rendle, Gantner, Freudenthaler, and
  Schmidt{-}Thieme}{Rendle et~al\mbox{.}}{2011}]%
        {FM}
\bibfield{author}{\bibinfo{person}{Steffen Rendle}, \bibinfo{person}{Zeno
  Gantner}, \bibinfo{person}{Christoph Freudenthaler}, {and}
  \bibinfo{person}{Lars Schmidt{-}Thieme}.} \bibinfo{year}{2011}\natexlab{}.
\newblock \showarticletitle{Fast context-aware recommendations with
  factorization machines}. In \bibinfo{booktitle}{\emph{{SIGIR}}}.
  \bibinfo{pages}{635--644}.
\newblock


\bibitem[\protect\citeauthoryear{Tay, Anh~Tuan, and Hui}{Tay
  et~al\mbox{.}}{2018}]%
        {tay2018latent}
\bibfield{author}{\bibinfo{person}{Yi Tay}, \bibinfo{person}{Luu Anh~Tuan},
  {and} \bibinfo{person}{Siu~Cheung Hui}.} \bibinfo{year}{2018}\natexlab{}.
\newblock \showarticletitle{Latent relational metric learning via memory-based
  attention for collaborative ranking}. In \bibinfo{booktitle}{\emph{WWW}}.
  \bibinfo{pages}{729--739}.
\newblock


\bibitem[\protect\citeauthoryear{van~den Berg, Kipf, and Welling}{van~den Berg
  et~al\mbox{.}}{2018}]%
        {GC-MC}
\bibfield{author}{\bibinfo{person}{Rianne van~den Berg},
  \bibinfo{person}{Thomas~N. Kipf}, {and} \bibinfo{person}{Max Welling}.}
  \bibinfo{year}{2018}\natexlab{}.
\newblock \showarticletitle{Graph Convolutional Matrix Completion}. In
  \bibinfo{booktitle}{\emph{KDD Workshop on Deep Learning Day}}.
\newblock


\bibitem[\protect\citeauthoryear{Velickovic, Cucurull, Casanova, Romero,
  Li{\`{o}}, and Bengio}{Velickovic et~al\mbox{.}}{2018}]%
        {GAT}
\bibfield{author}{\bibinfo{person}{Petar Velickovic}, \bibinfo{person}{Guillem
  Cucurull}, \bibinfo{person}{Arantxa Casanova}, \bibinfo{person}{Adriana
  Romero}, \bibinfo{person}{Pietro Li{\`{o}}}, {and} \bibinfo{person}{Yoshua
  Bengio}.} \bibinfo{year}{2018}\natexlab{}.
\newblock \showarticletitle{Graph Attention Networks}. In
  \bibinfo{booktitle}{\emph{{ICLR}}}.
\newblock


\bibitem[\protect\citeauthoryear{Wang, de~Vries, and Reinders}{Wang
  et~al\mbox{.}}{2006}]%
        {Wang:2006}
\bibfield{author}{\bibinfo{person}{Jun Wang}, \bibinfo{person}{Arjen~P. de
  Vries}, {and} \bibinfo{person}{Marcel J.~T. Reinders}.}
  \bibinfo{year}{2006}\natexlab{}.
\newblock \showarticletitle{Unifying User-based and Item-based Collaborative
  Filtering Approaches by Similarity Fusion}. In
  \bibinfo{booktitle}{\emph{SIGIR}}. \bibinfo{pages}{501--508}.
\newblock


\bibitem[\protect\citeauthoryear{Wang, He, Cao, Liu, and Chua}{Wang
  et~al\mbox{.}}{2019a}]%
        {KGAT}
\bibfield{author}{\bibinfo{person}{Xiang Wang}, \bibinfo{person}{Xiangnan He},
  \bibinfo{person}{Yixin Cao}, \bibinfo{person}{Meng Liu}, {and}
  \bibinfo{person}{Tat{-}Seng Chua}.} \bibinfo{year}{2019}\natexlab{a}.
\newblock \showarticletitle{KGAT: Knowledge Graph Attention Network for
  Recommendation}. In \bibinfo{booktitle}{\emph{{KDD}}}.
  \bibinfo{pages}{950--958}.
\newblock


\bibitem[\protect\citeauthoryear{Wang, He, Wang, Feng, and Chua}{Wang
  et~al\mbox{.}}{2019b}]%
        {NGCF}
\bibfield{author}{\bibinfo{person}{Xiang Wang}, \bibinfo{person}{Xiangnan He},
  \bibinfo{person}{Meng Wang}, \bibinfo{person}{Fuli Feng}, {and}
  \bibinfo{person}{Tat{-}Seng Chua}.} \bibinfo{year}{2019}\natexlab{b}.
\newblock \showarticletitle{Neural Graph Collaborative Filtering}. In
  \bibinfo{booktitle}{\emph{SIGIR}}. \bibinfo{pages}{165--174}.
\newblock


\bibitem[\protect\citeauthoryear{Wu, Jr., Zhang, Fifty, Yu, and Weinberger}{Wu
  et~al\mbox{.}}{2019a}]%
        {SGCN}
\bibfield{author}{\bibinfo{person}{Felix Wu}, \bibinfo{person}{Amauri H.~Souza
  Jr.}, \bibinfo{person}{Tianyi Zhang}, \bibinfo{person}{Christopher Fifty},
  \bibinfo{person}{Tao Yu}, {and} \bibinfo{person}{Kilian~Q. Weinberger}.}
  \bibinfo{year}{2019}\natexlab{a}.
\newblock \showarticletitle{Simplifying Graph Convolutional Networks}. In
  \bibinfo{booktitle}{\emph{ICML}}. \bibinfo{pages}{6861--6871}.
\newblock


\bibitem[\protect\citeauthoryear{Wu, Sun, Fu, Hong, Wang, and Wang}{Wu
  et~al\mbox{.}}{2019b}]%
        {GCNSocial}
\bibfield{author}{\bibinfo{person}{Le Wu}, \bibinfo{person}{Peijie Sun},
  \bibinfo{person}{Yanjie Fu}, \bibinfo{person}{Richang Hong},
  \bibinfo{person}{Xiting Wang}, {and} \bibinfo{person}{Meng Wang}.}
  \bibinfo{year}{2019}\natexlab{b}.
\newblock \showarticletitle{A Neural Influence Diffusion Model for Social
  Recommendation}. In \bibinfo{booktitle}{\emph{SIGIR}}.
  \bibinfo{pages}{235--244}.
\newblock


\bibitem[\protect\citeauthoryear{Xu, Hu, Leskovec, and Jegelka}{Xu
  et~al\mbox{.}}{2018}]%
        {GIN}
\bibfield{author}{\bibinfo{person}{Keyulu Xu}, \bibinfo{person}{Weihua Hu},
  \bibinfo{person}{Jure Leskovec}, {and} \bibinfo{person}{Stefanie Jegelka}.}
  \bibinfo{year}{2018}\natexlab{}.
\newblock \showarticletitle{How powerful are graph neural networks?}. In
  \bibinfo{booktitle}{\emph{{ICLR}}}.
\newblock


\bibitem[\protect\citeauthoryear{Yang, Chen, Wang, and Tsai}{Yang
  et~al\mbox{.}}{2018}]%
        {HOP-rec}
\bibfield{author}{\bibinfo{person}{Jheng{-}Hong Yang},
  \bibinfo{person}{Chih{-}Ming Chen}, \bibinfo{person}{Chuan{-}Ju Wang}, {and}
  \bibinfo{person}{Ming{-}Feng Tsai}.} \bibinfo{year}{2018}\natexlab{}.
\newblock \showarticletitle{HOP-rec: high-order proximity for implicit
  recommendation}. In \bibinfo{booktitle}{\emph{RecSys}}.
  \bibinfo{pages}{140--144}.
\newblock


\bibitem[\protect\citeauthoryear{Yin, Wang, Nie, He, Hong, and Chua}{Yin
  et~al\mbox{.}}{2019}]%
        {MMGCN}
\bibfield{author}{\bibinfo{person}{Yinwei Yin}, \bibinfo{person}{Xiang Wang},
  \bibinfo{person}{Liqiang Nie}, \bibinfo{person}{Xiangnan He},
  \bibinfo{person}{Richang Hong}, {and} \bibinfo{person}{Tat-Seng Chua}.}
  \bibinfo{year}{2019}\natexlab{}.
\newblock \showarticletitle{MMGCN: Multimodal Graph Convolution Network for
  Personalized Recommendation of Micro-video}. In
  \bibinfo{booktitle}{\emph{MM}}.
\newblock


\bibitem[\protect\citeauthoryear{Ying, He, Chen, Eksombatchai, Hamilton, and
  Leskovec}{Ying et~al\mbox{.}}{2018}]%
        {PinSage}
\bibfield{author}{\bibinfo{person}{Rex Ying}, \bibinfo{person}{Ruining He},
  \bibinfo{person}{Kaifeng Chen}, \bibinfo{person}{Pong Eksombatchai},
  \bibinfo{person}{William~L. Hamilton}, {and} \bibinfo{person}{Jure
  Leskovec}.} \bibinfo{year}{2018}\natexlab{}.
\newblock \showarticletitle{Graph Convolutional Neural Networks for Web-Scale
  Recommender Systems}. In \bibinfo{booktitle}{\emph{{KDD (Data Science
  track)}}}. \bibinfo{pages}{974--983}.
\newblock


\bibitem[\protect\citeauthoryear{Yuan, He, Karatzoglou, and Zhang}{Yuan
  et~al\mbox{.}}{2020}]%
        {PeterRec}
\bibfield{author}{\bibinfo{person}{Fajie Yuan}, \bibinfo{person}{Xiangnan He},
  \bibinfo{person}{Alexandros Karatzoglou}, {and} \bibinfo{person}{Liguang
  Zhang}.} \bibinfo{year}{2020}\natexlab{}.
\newblock \showarticletitle{Parameter-Efficient Transfer from Sequential
  Behaviors for User Modeling and Recommendation}. In
  \bibinfo{booktitle}{\emph{SIGIR}}.
\newblock


\bibitem[\protect\citeauthoryear{Zhao, Li, and Fu}{Zhao et~al\mbox{.}}{2019}]%
        {zhao2019cross}
\bibfield{author}{\bibinfo{person}{Cheng Zhao}, \bibinfo{person}{Chenliang Li},
  {and} \bibinfo{person}{Cong Fu}.} \bibinfo{year}{2019}\natexlab{}.
\newblock \showarticletitle{Cross-Domain Recommendation via Preference
  Propagation GraphNet}. In \bibinfo{booktitle}{\emph{CIKM}}.
  \bibinfo{pages}{2165--2168}.
\newblock


\bibitem[\protect\citeauthoryear{Zhu, Feng, He, Wang, Li, Zheng, and Zhang}{Zhu
  et~al\mbox{.}}{2020}]%
        {BGNN}
\bibfield{author}{\bibinfo{person}{Hongmin Zhu}, \bibinfo{person}{Fuli Feng},
  \bibinfo{person}{Xiangnan He}, \bibinfo{person}{Xiang Wang},
  \bibinfo{person}{Yan Li}, \bibinfo{person}{Kai Zheng}, {and}
  \bibinfo{person}{Yongdong Zhang}.} \bibinfo{year}{2020}\natexlab{}.
\newblock \showarticletitle{Bilinear Graph Neural Network with Neighbor
  Interactions}. In \bibinfo{booktitle}{\emph{IJCAI}}.
\newblock


\end{thebibliography}



\begin{thebibliography}{42}


\ifx \showCODEN    \undefined \def \showCODEN     #1{\unskip}     \fi
\ifx \showDOI      \undefined \def \showDOI       #1{#1}\fi
\ifx \showISBNx    \undefined \def \showISBNx     #1{\unskip}     \fi
\ifx \showISBNxiii \undefined \def \showISBNxiii  #1{\unskip}     \fi
\ifx \showISSN     \undefined \def \showISSN      #1{\unskip}     \fi
\ifx \showLCCN     \undefined \def \showLCCN      #1{\unskip}     \fi
\ifx \shownote     \undefined \def \shownote      #1{#1}          \fi
\ifx \showarticletitle \undefined \def \showarticletitle #1{#1}   \fi
\ifx \showURL      \undefined \def \showURL       {\relax}        \fi
\providecommand\bibfield[2]{#2}
\providecommand\bibinfo[2]{#2}
\providecommand\natexlab[1]{#1}
\providecommand\showeprint[2][]{arXiv:#2}

\bibitem[\protect\citeauthoryear{Cao, Wang, He, Hu, and Chua}{Cao
  et~al\mbox{.}}{2019}]%
        {KTUP}
\bibfield{author}{\bibinfo{person}{Yixin Cao}, \bibinfo{person}{Xiang Wang},
  \bibinfo{person}{Xiangnan He}, \bibinfo{person}{Zikun Hu}, {and}
  \bibinfo{person}{Tat{-}Seng Chua}.} \bibinfo{year}{2019}\natexlab{}.
\newblock \showarticletitle{Unifying Knowledge Graph Learning and
  Recommendation: Towards a Better Understanding of User Preferences}. In
  \bibinfo{booktitle}{\emph{{WWW}}}.
\newblock


\bibitem[\protect\citeauthoryear{Chen, Zhang, He, Nie, Liu, and Chua}{Chen
  et~al\mbox{.}}{2017}]%
        {ACF}
\bibfield{author}{\bibinfo{person}{Jingyuan Chen}, \bibinfo{person}{Hanwang
  Zhang}, \bibinfo{person}{Xiangnan He}, \bibinfo{person}{Liqiang Nie},
  \bibinfo{person}{Wei Liu}, {and} \bibinfo{person}{Tat{-}Seng Chua}.}
  \bibinfo{year}{2017}\natexlab{}.
\newblock \showarticletitle{Attentive Collaborative Filtering: Multimedia
  Recommendation with Item- and Component-Level Attention}. In
  \bibinfo{booktitle}{\emph{{SIGIR}}}. \bibinfo{pages}{335--344}.
\newblock


\bibitem[\protect\citeauthoryear{Cheng, Ding, Zhu, and Kankanhalli}{Cheng
  et~al\mbox{.}}{2018}]%
        {DBLP:conf/www/ChengDZK18}
\bibfield{author}{\bibinfo{person}{Zhiyong Cheng}, \bibinfo{person}{Ying Ding},
  \bibinfo{person}{Lei Zhu}, {and} \bibinfo{person}{Mohan~S. Kankanhalli}.}
  \bibinfo{year}{2018}\natexlab{}.
\newblock \showarticletitle{Aspect-Aware Latent Factor Model: Rating Prediction
  with Ratings and Reviews}. In \bibinfo{booktitle}{\emph{{WWW}}}.
  \bibinfo{pages}{639--648}.
\newblock


\bibitem[\protect\citeauthoryear{Defferrard, Bresson, and
  Vandergheynst}{Defferrard et~al\mbox{.}}{2016}]%
        {FirstGCN}
\bibfield{author}{\bibinfo{person}{Micha{\"{e}}l Defferrard},
  \bibinfo{person}{Xavier Bresson}, {and} \bibinfo{person}{Pierre
  Vandergheynst}.} \bibinfo{year}{2016}\natexlab{}.
\newblock \showarticletitle{Convolutional Neural Networks on Graphs with Fast
  Localized Spectral Filtering}. In \bibinfo{booktitle}{\emph{{NeurIPS}}}.
  \bibinfo{pages}{3837--3845}.
\newblock


\bibitem[\protect\citeauthoryear{Ebesu, Shen, and Fang}{Ebesu
  et~al\mbox{.}}{2018}]%
        {CMN}
\bibfield{author}{\bibinfo{person}{Travis Ebesu}, \bibinfo{person}{Bin Shen},
  {and} \bibinfo{person}{Yi Fang}.} \bibinfo{year}{2018}\natexlab{}.
\newblock \showarticletitle{Collaborative Memory Network for Recommendation
  Systems}. In \bibinfo{booktitle}{\emph{{SIGIR}}}. \bibinfo{pages}{515--524}.
\newblock


\bibitem[\protect\citeauthoryear{Glorot and Bengio}{Glorot and Bengio}{2010}]%
        {Xarvier}
\bibfield{author}{\bibinfo{person}{Xavier Glorot} {and} \bibinfo{person}{Yoshua
  Bengio}.} \bibinfo{year}{2010}\natexlab{}.
\newblock \showarticletitle{Understanding the difficulty of training deep
  feedforward neural networks}. In \bibinfo{booktitle}{\emph{{AISTATS}}}.
  \bibinfo{pages}{249--256}.
\newblock


\bibitem[\protect\citeauthoryear{Gori and Pucci}{Gori and Pucci}{2007}]%
        {ItemRank}
\bibfield{author}{\bibinfo{person}{Marco Gori} {and} \bibinfo{person}{Augusto
  Pucci}.} \bibinfo{year}{2007}\natexlab{}.
\newblock \showarticletitle{ItemRank: {A} Random-Walk Based Scoring Algorithm
  for Recommender Engines}. In \bibinfo{booktitle}{\emph{{IJCAI}}}.
  \bibinfo{pages}{2766--2771}.
\newblock


\bibitem[\protect\citeauthoryear{Hamilton, Ying, and Leskovec}{Hamilton
  et~al\mbox{.}}{2017}]%
        {GraphSAGE}
\bibfield{author}{\bibinfo{person}{William~L. Hamilton},
  \bibinfo{person}{Zhitao Ying}, {and} \bibinfo{person}{Jure Leskovec}.}
  \bibinfo{year}{2017}\natexlab{}.
\newblock \showarticletitle{Inductive Representation Learning on Large Graphs}.
  In \bibinfo{booktitle}{\emph{{NeurIPS}}}. \bibinfo{pages}{1025--1035}.
\newblock


\bibitem[\protect\citeauthoryear{He and McAuley}{He and McAuley}{2016a}]%
        {amazon-review}
\bibfield{author}{\bibinfo{person}{Ruining He} {and} \bibinfo{person}{Julian
  McAuley}.} \bibinfo{year}{2016}\natexlab{a}.
\newblock \showarticletitle{Ups and Downs: Modeling the Visual Evolution of
  Fashion Trends with One-Class Collaborative Filtering}. In
  \bibinfo{booktitle}{\emph{{WWW}}}. \bibinfo{pages}{507--517}.
\newblock


\bibitem[\protect\citeauthoryear{He and McAuley}{He and McAuley}{2016b}]%
        {VBPR}
\bibfield{author}{\bibinfo{person}{Ruining He} {and} \bibinfo{person}{Julian
  McAuley}.} \bibinfo{year}{2016}\natexlab{b}.
\newblock \showarticletitle{{VBPR:} Visual Bayesian Personalized Ranking from
  Implicit Feedback}. In \bibinfo{booktitle}{\emph{{AAAI}}}.
  \bibinfo{pages}{144--150}.
\newblock


\bibitem[\protect\citeauthoryear{He and Chua}{He and Chua}{2017}]%
        {NFM}
\bibfield{author}{\bibinfo{person}{Xiangnan He} {and}
  \bibinfo{person}{Tat{-}Seng Chua}.} \bibinfo{year}{2017}\natexlab{}.
\newblock \showarticletitle{Neural Factorization Machines for Sparse Predictive
  Analytics}. In \bibinfo{booktitle}{\emph{{SIGIR}}}.
  \bibinfo{pages}{355--364}.
\newblock


\bibitem[\protect\citeauthoryear{He, Gao, Kan, and Wang}{He
  et~al\mbox{.}}{2017a}]%
        {BiRank}
\bibfield{author}{\bibinfo{person}{Xiangnan He}, \bibinfo{person}{Ming Gao},
  \bibinfo{person}{Min{-}Yen Kan}, {and} \bibinfo{person}{Dingxian Wang}.}
  \bibinfo{year}{2017}\natexlab{a}.
\newblock \showarticletitle{BiRank: Towards Ranking on Bipartite Graphs}.
\newblock \bibinfo{journal}{\emph{TKDE}} \bibinfo{volume}{29},
  \bibinfo{number}{1} (\bibinfo{year}{2017}), \bibinfo{pages}{57--71}.
\newblock


\bibitem[\protect\citeauthoryear{He, He, Du, and Chua}{He
  et~al\mbox{.}}{2018}]%
        {APR}
\bibfield{author}{\bibinfo{person}{Xiangnan He}, \bibinfo{person}{Zhankui He},
  \bibinfo{person}{Xiaoyu Du}, {and} \bibinfo{person}{Tat{-}Seng Chua}.}
  \bibinfo{year}{2018}\natexlab{}.
\newblock \showarticletitle{Adversarial Personalized Ranking for
  Recommendation}. In \bibinfo{booktitle}{\emph{{SIGIR}}}.
  \bibinfo{pages}{355--364}.
\newblock


\bibitem[\protect\citeauthoryear{He, Liao, Zhang, Nie, Hu, and Chua}{He
  et~al\mbox{.}}{2017b}]%
        {NCF}
\bibfield{author}{\bibinfo{person}{Xiangnan He}, \bibinfo{person}{Lizi Liao},
  \bibinfo{person}{Hanwang Zhang}, \bibinfo{person}{Liqiang Nie},
  \bibinfo{person}{Xia Hu}, {and} \bibinfo{person}{Tat{-}Seng Chua}.}
  \bibinfo{year}{2017}\natexlab{b}.
\newblock \showarticletitle{Neural Collaborative Filtering}. In
  \bibinfo{booktitle}{\emph{{WWW}}}. \bibinfo{pages}{173--182}.
\newblock


\bibitem[\protect\citeauthoryear{Hsieh, Yang, Cui, Lin, Belongie, and
  Estrin}{Hsieh et~al\mbox{.}}{2017}]%
        {CML}
\bibfield{author}{\bibinfo{person}{Cheng{-}Kang Hsieh}, \bibinfo{person}{Longqi
  Yang}, \bibinfo{person}{Yin Cui}, \bibinfo{person}{Tsung{-}Yi Lin},
  \bibinfo{person}{Serge~J. Belongie}, {and} \bibinfo{person}{Deborah Estrin}.}
  \bibinfo{year}{2017}\natexlab{}.
\newblock \showarticletitle{Collaborative Metric Learning}. In
  \bibinfo{booktitle}{\emph{{WWW}}}. \bibinfo{pages}{193--201}.
\newblock


\bibitem[\protect\citeauthoryear{Kabbur, Ning, and Karypis}{Kabbur
  et~al\mbox{.}}{2013}]%
        {FISM}
\bibfield{author}{\bibinfo{person}{Santosh Kabbur}, \bibinfo{person}{Xia Ning},
  {and} \bibinfo{person}{George Karypis}.} \bibinfo{year}{2013}\natexlab{}.
\newblock \showarticletitle{{FISM:} factored item similarity models for top-N
  recommender systems}. In \bibinfo{booktitle}{\emph{{KDD}}}.
  \bibinfo{pages}{659--667}.
\newblock


\bibitem[\protect\citeauthoryear{Kingma and Ba}{Kingma and Ba}{2015}]%
        {Adam}
\bibfield{author}{\bibinfo{person}{Diederik~P. Kingma} {and}
  \bibinfo{person}{Jimmy Ba}.} \bibinfo{year}{2015}\natexlab{}.
\newblock \showarticletitle{Adam: {A} Method for Stochastic Optimization}. In
  \bibinfo{booktitle}{\emph{{ICLR}}}.
\newblock


\bibitem[\protect\citeauthoryear{Kipf and Welling}{Kipf and Welling}{2017}]%
        {GCN}
\bibfield{author}{\bibinfo{person}{Thomas~N. Kipf} {and} \bibinfo{person}{Max
  Welling}.} \bibinfo{year}{2017}\natexlab{}.
\newblock \showarticletitle{Semi-Supervised Classification with Graph
  Convolutional Networks}. In \bibinfo{booktitle}{\emph{{ICLR}}}.
\newblock


\bibitem[\protect\citeauthoryear{Koren}{Koren}{2008}]%
        {SVDFeatures}
\bibfield{author}{\bibinfo{person}{Yehuda Koren}.}
  \bibinfo{year}{2008}\natexlab{}.
\newblock \showarticletitle{Factorization meets the neighborhood: a
  multifaceted collaborative filtering model}. In
  \bibinfo{booktitle}{\emph{{KDD}}}. \bibinfo{pages}{426--434}.
\newblock


\bibitem[\protect\citeauthoryear{Koren, Bell, and Volinsky}{Koren
  et~al\mbox{.}}{2009}]%
        {MF}
\bibfield{author}{\bibinfo{person}{Yehuda Koren}, \bibinfo{person}{Robert~M.
  Bell}, {and} \bibinfo{person}{Chris Volinsky}.}
  \bibinfo{year}{2009}\natexlab{}.
\newblock \showarticletitle{Matrix Factorization Techniques for Recommender
  Systems}.
\newblock \bibinfo{journal}{\emph{{IEEE} Computer}} \bibinfo{volume}{42},
  \bibinfo{number}{8} (\bibinfo{year}{2009}), \bibinfo{pages}{30--37}.
\newblock


\bibitem[\protect\citeauthoryear{Liang, Charlin, McInerney, and Blei}{Liang
  et~al\mbox{.}}{2016}]%
        {gowalla}
\bibfield{author}{\bibinfo{person}{Dawen Liang}, \bibinfo{person}{Laurent
  Charlin}, \bibinfo{person}{James McInerney}, {and} \bibinfo{person}{David~M.
  Blei}.} \bibinfo{year}{2016}\natexlab{}.
\newblock \showarticletitle{Modeling User Exposure in Recommendation}. In
  \bibinfo{booktitle}{\emph{{WWW}}}. \bibinfo{pages}{951--961}.
\newblock


\bibitem[\protect\citeauthoryear{Liu, Wang, Zhang, Shah, Xia, Yang, and Li}{Liu
  et~al\mbox{.}}{2017}]%
        {DBLP:conf/mm/LiuWZSXYL17}
\bibfield{author}{\bibinfo{person}{Zhenguang Liu}, \bibinfo{person}{Zepeng
  Wang}, \bibinfo{person}{Luming Zhang}, \bibinfo{person}{Rajiv~Ratn Shah},
  \bibinfo{person}{Yingjie Xia}, \bibinfo{person}{Yi Yang}, {and}
  \bibinfo{person}{Xuelong Li}.} \bibinfo{year}{2017}\natexlab{}.
\newblock \showarticletitle{FastShrinkage: Perceptually-aware Retargeting
  Toward Mobile Platforms}. In \bibinfo{booktitle}{\emph{{MM}}}.
  \bibinfo{pages}{501--509}.
\newblock


\bibitem[\protect\citeauthoryear{Maas, Hannun, and Ng}{Maas
  et~al\mbox{.}}{2013}]%
        {LeakyRelu}
\bibfield{author}{\bibinfo{person}{Andrew~L Maas}, \bibinfo{person}{Awni~Y
  Hannun}, {and} \bibinfo{person}{Andrew~Y Ng}.}
  \bibinfo{year}{2013}\natexlab{}.
\newblock \showarticletitle{Rectifier nonlinearities improve neural network
  acoustic models}. In \bibinfo{booktitle}{\emph{ICML}}.
\newblock


\bibitem[\protect\citeauthoryear{Nikolakopoulos and Karypis}{Nikolakopoulos and
  Karypis}{2019}]%
        {RecWalk}
\bibfield{author}{\bibinfo{person}{Athanasios~N Nikolakopoulos} {and}
  \bibinfo{person}{George Karypis}.} \bibinfo{year}{2019}\natexlab{}.
\newblock \showarticletitle{RecWalk: Nearly Uncoupled Random Walks for Top-N
  Recommendation}.
\newblock  (\bibinfo{year}{2019}).
\newblock


\bibitem[\protect\citeauthoryear{Qiu, Tang, Ma, Dong, Wang, and Tang}{Qiu
  et~al\mbox{.}}{2018}]%
        {DeepInf}
\bibfield{author}{\bibinfo{person}{Jiezhong Qiu}, \bibinfo{person}{Jian Tang},
  \bibinfo{person}{Hao Ma}, \bibinfo{person}{Yuxiao Dong},
  \bibinfo{person}{Kuansan Wang}, {and} \bibinfo{person}{Jie Tang}.}
  \bibinfo{year}{2018}\natexlab{}.
\newblock \showarticletitle{DeepInf: Social Influence Prediction with Deep
  Learning}. In \bibinfo{booktitle}{\emph{{KDD}}}. \bibinfo{pages}{2110--2119}.
\newblock


\bibitem[\protect\citeauthoryear{Rendle, Freudenthaler, Gantner, and
  Schmidt{-}Thieme}{Rendle et~al\mbox{.}}{2009}]%
        {BPRMF}
\bibfield{author}{\bibinfo{person}{Steffen Rendle}, \bibinfo{person}{Christoph
  Freudenthaler}, \bibinfo{person}{Zeno Gantner}, {and} \bibinfo{person}{Lars
  Schmidt{-}Thieme}.} \bibinfo{year}{2009}\natexlab{}.
\newblock \showarticletitle{{BPR:} Bayesian Personalized Ranking from Implicit
  Feedback}. In \bibinfo{booktitle}{\emph{{UAI}}}. \bibinfo{pages}{452--461}.
\newblock


\bibitem[\protect\citeauthoryear{Song, Feng, Han, Yang, Liu, and Nie}{Song
  et~al\mbox{.}}{2018}]%
        {SongFHYLN18}
\bibfield{author}{\bibinfo{person}{Xuemeng Song}, \bibinfo{person}{Fuli Feng},
  \bibinfo{person}{Xianjing Han}, \bibinfo{person}{Xin Yang},
  \bibinfo{person}{Wei Liu}, {and} \bibinfo{person}{Liqiang Nie}.}
  \bibinfo{year}{2018}\natexlab{}.
\newblock \showarticletitle{Neural Compatibility Modeling with Attentive
  Knowledge Distillation}. In \bibinfo{booktitle}{\emph{SIGIR}}.
  \bibinfo{pages}{5--14}.
\newblock


\bibitem[\protect\citeauthoryear{Tay, Anh~Tuan, and Hui}{Tay
  et~al\mbox{.}}{2018}]%
        {tay2018latent}
\bibfield{author}{\bibinfo{person}{Yi Tay}, \bibinfo{person}{Luu Anh~Tuan},
  {and} \bibinfo{person}{Siu~Cheung Hui}.} \bibinfo{year}{2018}\natexlab{}.
\newblock \showarticletitle{Latent relational metric learning via memory-based
  attention for collaborative ranking}. In \bibinfo{booktitle}{\emph{WWW}}.
  \bibinfo{pages}{729--739}.
\newblock


\bibitem[\protect\citeauthoryear{van~den Berg, Kipf, and Welling}{van~den Berg
  et~al\mbox{.}}{2017}]%
        {GC-MC}
\bibfield{author}{\bibinfo{person}{Rianne van~den Berg},
  \bibinfo{person}{Thomas~N. Kipf}, {and} \bibinfo{person}{Max Welling}.}
  \bibinfo{year}{2017}\natexlab{}.
\newblock \showarticletitle{Graph Convolutional Matrix Completion}. In
  \bibinfo{booktitle}{\emph{{KDD}}}.
\newblock


\bibitem[\protect\citeauthoryear{Wang, Wang, and Yeung}{Wang
  et~al\mbox{.}}{2015}]%
        {CDL}
\bibfield{author}{\bibinfo{person}{Hao Wang}, \bibinfo{person}{Naiyan Wang},
  {and} \bibinfo{person}{Dit{-}Yan Yeung}.} \bibinfo{year}{2015}\natexlab{}.
\newblock \showarticletitle{Collaborative Deep Learning for Recommender
  Systems}. In \bibinfo{booktitle}{\emph{{KDD}}}. \bibinfo{pages}{1235--1244}.
\newblock


\bibitem[\protect\citeauthoryear{Wang, He, Cao, Liu, and Chua}{Wang
  et~al\mbox{.}}{2019a}]%
        {KGAT}
\bibfield{author}{\bibinfo{person}{Xiang Wang}, \bibinfo{person}{Xiangnan He},
  \bibinfo{person}{Yixin Cao}, \bibinfo{person}{Meng Liu}, {and}
  \bibinfo{person}{Tat{-}Seng Chua}.} \bibinfo{year}{2019}\natexlab{a}.
\newblock \showarticletitle{KGAT: Knowledge Graph Attention Network for
  Recommendation}. In \bibinfo{booktitle}{\emph{{KDD}}}.
\newblock


\bibitem[\protect\citeauthoryear{Wang, He, Feng, Nie, and Chua}{Wang
  et~al\mbox{.}}{2018}]%
        {TEM}
\bibfield{author}{\bibinfo{person}{Xiang Wang}, \bibinfo{person}{Xiangnan He},
  \bibinfo{person}{Fuli Feng}, \bibinfo{person}{Liqiang Nie}, {and}
  \bibinfo{person}{Tat{-}Seng Chua}.} \bibinfo{year}{2018}\natexlab{}.
\newblock \showarticletitle{{TEM:} Tree-enhanced Embedding Model for
  Explainable Recommendation}. In \bibinfo{booktitle}{\emph{{WWW}}}.
  \bibinfo{pages}{1543--1552}.
\newblock


\bibitem[\protect\citeauthoryear{Wang, He, Nie, and Chua}{Wang
  et~al\mbox{.}}{2017}]%
        {ItemSilk}
\bibfield{author}{\bibinfo{person}{Xiang Wang}, \bibinfo{person}{Xiangnan He},
  \bibinfo{person}{Liqiang Nie}, {and} \bibinfo{person}{Tat{-}Seng Chua}.}
  \bibinfo{year}{2017}\natexlab{}.
\newblock \showarticletitle{Item Silk Road: Recommending Items from Information
  Domains to Social Users}. In \bibinfo{booktitle}{\emph{{SIGIR}}}.
  \bibinfo{pages}{185--194}.
\newblock


\bibitem[\protect\citeauthoryear{Wang, Wang, Xu, He, Cao, and Chua}{Wang
  et~al\mbox{.}}{2019b}]%
        {KPRN}
\bibfield{author}{\bibinfo{person}{Xiang Wang}, \bibinfo{person}{Dingxian
  Wang}, \bibinfo{person}{Canran Xu}, \bibinfo{person}{Xiangnan He},
  \bibinfo{person}{Yixin Cao}, {and} \bibinfo{person}{Tat{-}Seng Chua}.}
  \bibinfo{year}{2019}\natexlab{b}.
\newblock \showarticletitle{Explainable Reasoning over Knowledge Graphs for
  Recommendation}. In \bibinfo{booktitle}{\emph{{AAAI}}}.
\newblock


\bibitem[\protect\citeauthoryear{Wu, DuBois, Zheng, and Ester}{Wu
  et~al\mbox{.}}{2016}]%
        {AutoCF}
\bibfield{author}{\bibinfo{person}{Yao Wu}, \bibinfo{person}{Christopher
  DuBois}, \bibinfo{person}{Alice~X. Zheng}, {and} \bibinfo{person}{Martin
  Ester}.} \bibinfo{year}{2016}\natexlab{}.
\newblock \showarticletitle{Collaborative Denoising Auto-Encoders for Top-N
  Recommender Systems}. In \bibinfo{booktitle}{\emph{{WSDM}}}.
  \bibinfo{pages}{153--162}.
\newblock


\bibitem[\protect\citeauthoryear{Xin, He, Zhang, Zhang, and Jose}{Xin
  et~al\mbox{.}}{2019}]%
        {RCF}
\bibfield{author}{\bibinfo{person}{Xin Xin}, \bibinfo{person}{Xiangnan He},
  \bibinfo{person}{Yongfeng Zhang}, \bibinfo{person}{Yongdong Zhang}, {and}
  \bibinfo{person}{Joemon Jose}.} \bibinfo{year}{2019}\natexlab{}.
\newblock \showarticletitle{Relational Collaborative Filtering:Modeling
  Multiple Item Relations for Recommendation}. In
  \bibinfo{booktitle}{\emph{{SIGIR}}}.
\newblock


\bibitem[\protect\citeauthoryear{Xu, Li, Tian, Sonobe, Kawarabayashi, and
  Jegelka}{Xu et~al\mbox{.}}{2018}]%
        {JumpKG}
\bibfield{author}{\bibinfo{person}{Keyulu Xu}, \bibinfo{person}{Chengtao Li},
  \bibinfo{person}{Yonglong Tian}, \bibinfo{person}{Tomohiro Sonobe},
  \bibinfo{person}{Ken{-}ichi Kawarabayashi}, {and} \bibinfo{person}{Stefanie
  Jegelka}.} \bibinfo{year}{2018}\natexlab{}.
\newblock \showarticletitle{Representation Learning on Graphs with Jumping
  Knowledge Networks}. In \bibinfo{booktitle}{\emph{{ICML}}},
  Vol.~\bibinfo{volume}{80}. \bibinfo{pages}{5449--5458}.
\newblock


\bibitem[\protect\citeauthoryear{Xue, He, Wang, Xu, Liu, and Hong}{Xue
  et~al\mbox{.}}{2019}]%
        {DeepICF}
\bibfield{author}{\bibinfo{person}{Feng Xue}, \bibinfo{person}{Xiangnan He},
  \bibinfo{person}{Xiang Wang}, \bibinfo{person}{Jiandong Xu},
  \bibinfo{person}{Kai Liu}, {and} \bibinfo{person}{Richang Hong}.}
  \bibinfo{year}{2019}\natexlab{}.
\newblock \showarticletitle{Deep Item-based Collaborative Filtering for Top-N
  Recommendation}.
\newblock \bibinfo{journal}{\emph{TOIS}} \bibinfo{volume}{37},
  \bibinfo{number}{3} (\bibinfo{year}{2019}), \bibinfo{pages}{33:1--33:25}.
\newblock


\bibitem[\protect\citeauthoryear{Yang, Chen, Wang, and Tsai}{Yang
  et~al\mbox{.}}{2018}]%
        {HOP-rec}
\bibfield{author}{\bibinfo{person}{Jheng{-}Hong Yang},
  \bibinfo{person}{Chih{-}Ming Chen}, \bibinfo{person}{Chuan{-}Ju Wang}, {and}
  \bibinfo{person}{Ming{-}Feng Tsai}.} \bibinfo{year}{2018}\natexlab{}.
\newblock \showarticletitle{HOP-rec: high-order proximity for implicit
  recommendation}. In \bibinfo{booktitle}{\emph{RecSys}}.
  \bibinfo{pages}{140--144}.
\newblock


\bibitem[\protect\citeauthoryear{Yang, He, Wang, Ma, Feng, Wang, and Chua}{Yang
  et~al\mbox{.}}{2019}]%
        {FashionInterpretable}
\bibfield{author}{\bibinfo{person}{Xun Yang}, \bibinfo{person}{Xiangnan He},
  \bibinfo{person}{Xiang Wang}, \bibinfo{person}{Yunshan Ma},
  \bibinfo{person}{Fuli Feng}, \bibinfo{person}{Meng Wang}, {and}
  \bibinfo{person}{Tat-Seng Chua}.} \bibinfo{year}{2019}\natexlab{}.
\newblock \showarticletitle{Interpretable Fashion Matching with Rich
  Attributes}. In \bibinfo{booktitle}{\emph{{SIGIR}}}.
\newblock


\bibitem[\protect\citeauthoryear{Ying, He, Chen, Eksombatchai, Hamilton, and
  Leskovec}{Ying et~al\mbox{.}}{2018}]%
        {PinSage}
\bibfield{author}{\bibinfo{person}{Rex Ying}, \bibinfo{person}{Ruining He},
  \bibinfo{person}{Kaifeng Chen}, \bibinfo{person}{Pong Eksombatchai},
  \bibinfo{person}{William~L. Hamilton}, {and} \bibinfo{person}{Jure
  Leskovec}.} \bibinfo{year}{2018}\natexlab{}.
\newblock \showarticletitle{Graph Convolutional Neural Networks for Web-Scale
  Recommender Systems}. In \bibinfo{booktitle}{\emph{{KDD (Data Science
  track)}}}. \bibinfo{pages}{974--983}.
\newblock


\bibitem[\protect\citeauthoryear{Zheng, Lu, Jiang, Zhang, and Yu}{Zheng
  et~al\mbox{.}}{2018}]%
        {SpectralCF}
\bibfield{author}{\bibinfo{person}{Lei Zheng}, \bibinfo{person}{Chun{-}Ta Lu},
  \bibinfo{person}{Fei Jiang}, \bibinfo{person}{Jiawei Zhang}, {and}
  \bibinfo{person}{Philip~S. Yu}.} \bibinfo{year}{2018}\natexlab{}.
\newblock \showarticletitle{Spectral collaborative filtering}. In
  \bibinfo{booktitle}{\emph{RecSys}}. \bibinfo{pages}{311--319}.
\newblock


\end{thebibliography}
\balance
\scriptsize

\end{document}